\documentclass[fleqn,usenatbib,useAMS]{mnras}
\pdfoutput=1

\usepackage{newtxtext,newtxmath}

\usepackage[T1]{fontenc}

\DeclareRobustCommand{\VAN}[3]{#2}
\let\VANthebibliography\thebibliography
\def\thebibliography{\DeclareRobustCommand{\VAN}[3]{##3}\VANthebibliography}

\usepackage{subfig}

\usepackage{graphicx}	
\usepackage{amsmath}	
\usepackage{amssymb}	
\usepackage{natbib}
\usepackage{graphicx}

\title[Clustering-informed Cinematic Astrophysical Data Visualization]{Clustering-informed Cinematic Astrophysical Data Visualization with Application to the Moon-forming Terrestrial Synestia}

\author[Patrick D. Aleo et al.]{
Patrick D. Aleo,$^{1,2,3}$\thanks{E-mail: paleo2@illinois.edu}
Simon J. Lock$^{4}$,
Donna J. Cox,$^{2}$
Stuart A. Levy,$^{2}$
J. P. Naiman,$^{2}$\newauthor
A. J. Christensen,$^{2}$
Kalina Borkiewicz,$^{2}$
and Robert Patterson$^{2}$
\\
$^{1}$Department of Astronomy, University of Illinois at Urbana-Champaign, 
1002 West Green Street, 
Urbana, IL 61801, USA\\
$^{2}$Advanced Visualization Laboratory, National Center for Supercomputing Applications, 
1205 West Clark Street,
Urbana, IL 61801, USA\\
$^{3}$Fiddler Innovation Fellow\\
$^{4}$Division of Geological and Planetary Sciences, California Institute of Technology, 
1200 East California Boulevard,
Pasadena, CA 91125, USA\\
}

\date{Accepted XXX. Received YYY; in original form ZZZ}

\pubyear{2020}

\begin{document}
\label{firstpage}
\pagerange{\pageref{firstpage}--\pageref{lastpage}}
\maketitle

\begin{abstract}

Scientific visualization tools are currently not optimized to create cinematic, production-quality representations of numerical data for the purpose of science communication. In our pipeline \texttt{Estra}, we outline a step-by-step process from a raw simulation into a finished render as a way to teach non-experts in the field of visualization how to achieve production-quality outputs on their own. We demonstrate feasibility of using the visual effects software Houdini for cinematic astrophysical data visualization, informed by machine learning clustering algorithms. To demonstrate the capabilities of this pipeline, we used a post-impact, thermally-equilibrated Moon-forming synestia from \cite{Lock18}. Our approach aims to identify ``physically interpretable'' clusters, where clusters identified in an appropriate phase space (e.g. here we use a temperature-entropy phase-space) correspond to physically meaningful structures within the simulation data. Clustering results can then be used to highlight these structures by informing the color-mapping process in a simplified Houdini software shading network, where dissimilar phase-space clusters are mapped to different color values for easier visual identification. Cluster information can also be used in 3D position space, via Houdini's Scene View, to aid in physical cluster finding, simulation prototyping, and data exploration. Our clustering-based renders are compared to those created by the Advanced Visualization Lab (AVL) team for the full dome show ``Imagine the Moon'' as proof of concept. With \texttt{Estra}, scientists have a tool to create their own production-quality, data-driven visualizations.

\end{abstract}




\section{Introduction} \label{sec:intro}

Data visualization, the graphical display of either spatial or temporal data, is a wide field used for data analysis and communication \citep[\& references therein]{Borkiewicz19a, Borkiewicz19b, Punzo15, Hassan11, Barnes08, Price07}. The style and content of data visualization may vary \citep{Borkiewicz19a, Goodman12}, but there are generally three distinct paradigms: 
\begin{enumerate}
    \item ``Information visualization''---typically a two-dimensional representation of relational/non-spatial data via networks, graphs, and charts;
    \item  ``Traditional scientific visualization''---imagery created for three-dimensional spatial data to be analyzed predominantly by scientists for publications in peer-reviewed journals;
    \item ``Cinematic scientific visualization''---production quality, data-driven imagery with aesthetic appeal designed for mass audiences and large-format screens. 
\end{enumerate}

In cinematic scientific visualization, Hollywood techniques of composition, rendering quality, and camera design are chosen to bring about visually attractive presentations of the science. There is supporting evidence such attractive presentations are more educational than unattractive ones \citep{Cawthon07}, and spur interest in scientific topics, even those widely-considered monotonous or difficult-to-learn \citep{Arroio10, Dubeck94}. Because of cinematic scientific visualization's power to simultaneously entertain, educate, and provide new insight about the science in question, it is important that steps be taken to increase ubiquity and ease of use by non-visualization designers, like domain scientists. \par

By equipping scientists with the means and tools to create cinematic-driven visualizations of their own data, scientists can create new, insightful, broad-reaching imagery for use in all forms of science communication: from peer-reviewed publications, conferences, presentations, and simulation prototyping to public outreach, news media, interviews, and social media \citep{Borkiewicz19a, Shih19}. In an age of misinformation \citep{Borkiewicz17}, it has never been more important for data-driven visualizations to spearhead scientific communications directly from scientists themselves, even more so than popular science communicators. In fact, it has been shown that user-generated science content via scientists are more popular on Youtube than those created by professional scientific communicators \citep{Welbourne16}. \par

Currently, the niche field of cinematic scientific visualization is dominated by visualization and visual effects designers who, in communication with scientists, create accurate representations of the data. Scientists, particularly astrophysicists for this work, are largely focused on traditional scientific visualization, designed to be shared and understood by peers exclusively. According to \cite{Hassan11}, the astronomical community has put in ``limited effort'' to develop general purpose, 3D-visualization tools equipped to handle astronomical data. However, recently there has been some development in creating tools for such cinematic visualization. For example: tools implementing astronomical Adaptive Mesh Refinement (AMR) \citep{Berger84, Kahler02} data into the visual effects software Houdini via \texttt{ytini}\footnote{\url{http://ytini.com}} \citep{Naiman17, Borkiewicz19b}; generating physically-accurate artistic models of astronomical objects with 3D modeling software Blender \citep{Kent13, Kent15}; combining the analysis tools of \texttt{yt} \citep{Turk11} with  Blender via \texttt{AstroBlend} \citep{Naiman16}; and applying 3D interactive visualization software to astronomical FITS files via \texttt{SlicerAstro} \citep{Punzo17}. However, tools for scientists to develop cinematic visualizations remain limited and their use is not widespread. With the beginning of the ``Petascale Astronomy Era'' \citep{Hassan11} of next-generation sky surveys and supercomputing facilities, such developments have never been more crucial. \par

The computational cost of cinematic astrophysical data visualization is decreasing with improving technologies and techniques. As such, there will be a growing interdisciplinary environment for collaborations between scientists and digital artists \citep{Cox87}. Visual effects and modeling software like Houdini, Blender, and Unity can read-in scientific data with plug-ins and packages, and are waiting to be more widely-used by the astronomical community. This work will help scientists become their own storytellers and develop simple, yet effective cinematic astrophysical data visualizations using machine learning clustering algorithms and the visual effects software Houdini. \par


Machine learning and deep learning techniques and ideas have become prominent in astronomy because of their innate capability to process and analyze big data \citep[\& references therein]{Bell10, Pesenson10, Burke19, Baron19}. With the substantial amount of data generated from large current and future astronomical surveys and simulations, it is imperative that the field of cinematic astrophysical visualization not be left behind, and take advantage of the cutting edge software and hardware available. \par
 
The application of machine learning in visualization is a relatively new and underdeveloped field \citep{Ma07}. As we show in this work, one application of machine learning that is well suited for visualization is cluster finding. Machine learning algorithms can be used to discover interesting features and classify clusters, and in scientific studies it is up to the researcher to determine the context. Likewise, a visualization artist has a narrative or educational insight they wish to convey to an audience, and this decision informs what features the visualization will highlight. \par 

Often in visualization, only a small subset of all the data attributes (that is, the span of the $n$-dimensional dataset) are chosen for the final render due to computational costs, aesthetic quality, and other subjective reasoning such as human perception. The visualization artist will spend weeks implementing a ``guess and check'' method to see what ``looks best'' when it comes to lighting and coloring (shading) the simulation and adjusting which variables are mapped to luminance and opacity, etc. However, this can lead to key features or structures being overlooked or not emphasized with the appropriate importance. Clustering methods can inform/automate visualization decisions/processes, ensuring that the attributes chosen to be highlighted best represent the dataset. Automatically-generated colormaps are created based on the clustering results, and these features are highlighted in the render. Thus, physical structures within the dataset are emphasized. Additionally, clusters can identify small features/substructures that may be easily passed over otherwise. It is important that the clusters identified by the clustering algorithms are ``physically interpretable'' \citep{Milosavljevic18}, in that each cluster in some 2D phase-space corresponds to a physical structure in the 3D (spatial) dataset. This is useful for scientists to better understand relevant physical processes and enables them to investigate their data in a new way. \par
 
Here we outline a new pipeline for cinematic visualization of scientific data, \texttt{Estra}, that takes advantage of modern clustering algorithms and applies it to an example simulation of a Moon-forming giant impact \citep{Lock18}. \texttt{Estra} is advantageous to both visualization designers and domain scientists. It is a good starting point for a visualization designer because it reduces time spent on manual data exploration and previsualization. Moreover, it enables domain scientists to explore their data in new ways, and easily create high-fidelity scientific visualizations. \par
 
Section~\S \ref{sec:synestia} describes the post-impact synestia from \cite{Lock18} used as the example dataset. Section~\S \ref{sec:Estra} briefly discusses the \texttt{Estra} Python workflow, with a full step-by-step process outlined in the accompanying Python notebooks, and gives a theoretical overview of the clustering algorithm used in this work. Section~\S \ref{sec:Methods} outlines the procedure, from clustering the data and using its results, and describes simple assumptions that we used to build and inform a shader\footnote{A program that determines how 3D surface properties (lighting, color, etc.) of objects are rendered for each pixel.} network. Section~\S \ref{sec:Results} displays various final renders, and demonstrates the quality and validity of the visualizations. Lastly, we conclude in Section~\S \ref{sec:Conclusion}. The appendices include a short mathematical treatment of other popular clustering algorithms, as well as additional renders using the \texttt{Estra} shader with a perceptually-uniform emissive colormap. \par

Our code is publicly available at
\url{https://github.com/patrickaleo/estra}. \par

\section{The Moon-forming Terrestrial Synestia}\label{sec:synestia}



As an example dataset to demonstrate our pipeline, we use the output of a smoothed-particle hydrodynamics (SPH) \citep{Gingold77, Price07} simulation of a Moon-forming giant impact from \cite{Lock18}. SPH is a Lagrangian fluid dynamics method where the fluid is divided into particles with the dynamics of each particle governed by interactions with its nearest neighbors. The output is in the form of the properties (spatial position, velocity, thermodynamic properties, gravitational potential etc.) for each particle. Before pre-processing, there are 100,989 particles in our example simulation. \par

Giant impacts are collisions between planet-sized bodies which are common in the formation of our solar system and exosystems \citep{Raymond18}, and it is thought that the last impact the Earth experienced ejected sufficient material into orbit to form our Moon \citep{Cameron76, Hartmann75}. The Moon-forming giant impact was a highly energetic event and left the post-impact body rotating rapidly and with a silicate mantle that was substantially supercritical or vaporized \citep{Lock18, LockStewart17, Nakajima15}. \cite{LockStewart17} recently demonstrated that a subset of impacts are sufficiently energetic, and have high enough post-impact angular momentum to produce a previously unrecognized type of planetary structure: a synestia. Synestias could provide a new environment for the formation of the Moon \citep{Lock18} and are a topic of ongoing research. \par

For this work we considered the final time step of an impact that produced a Moon-forming synestia from \cite{Lock18}. In this example, the synestia was formed by a 0.1 $M_{Earth}$ body striking a 0.99 $M_{Earth}$ body spinning with a 2.3 hr period (an angular momentum of three times the present-day Earth-Moon angular momentum) at 15 km s$^{-1}$ and an impact parameter of 0.4. The simulation was run for 48 hours of simulation time, when the structure was nearly axisymmetric and had reached a quasi-hydrostatic equilibrium. The output was post-processed to simulate thermal equilibration after the impact, driven by processes that are not captured in the code (see supporting information of \cite{Lock18} for more details). The outer regions of the synestia were thermally equilibrated and a portion of the outer regions of the body were prescribed to be isentropic. Any condensed material was removed from the simulation and the remaining mass of multiphase particles was forced to lie on the vapor side of the liquid-vapor phase boundary. \par

The SPH simulation we use has previously been visualized by the Advanced Visualization Lab (AVL) at the National Center for Supercomputing Applications (NCSA) to produce part of an animation for the dome show ``Imagine the Moon''.\footnote{\url{https://www.adlerplanetarium.org/event/imagine-the-moon/}; an excerpt video can be viewed here: \url{https://www.youtube.com/watch?v=7e_6oyROHCU}} Using the same visual effects software Houdini to visualize the dataset, this work directly compares a machine learning clustering-informed cinematic visualization to a custom AVL cinematic visualization.

The synestia dataset in .csv format is available on the \texttt{Estra} Github page\footnote{\url{https://github.com/patrickaleo/estra}}, for easy replication of our results. \par

\section{\texttt{Estra} and clustering algorithms}\label{sec:Estra}

In this section we briefly discuss the \texttt{Estra} Python workflow, with a full step-by-step process outlined in the accompanying Python notebooks, as well as introduce Gaussian Mixture Model (GMM) theory.

\subsection{Outline of \texttt{Estra} integration with visual effects software Houdini}

Once the simulation data is fully loaded into Houdini\footnote{\url{www.sidefx.com}}, we can extract all attribute (parameter) values for each particle in the simulation output as a .csv file using a simple custom script in a Python Script\footnote{\url{https://www.sidefx.com/docs/houdini/nodes/obj/pythonscript.html}} object node. Alternatively, one can extract attribute values from the data file directly. This step is necessary because the clustering algorithms that use the attribute data cannot be performed in Houdini itself. We read in the .csv file into a Jupyter notebook and performed the clustering algorithms utilizing the \texttt{sklearn} Python package, and transferred our results back into Houdini to inform some visualization decisions, such as automating a clustering-based color temperature ramp in the material shader. This process is implemented in ``Estra.ipynb'', and is also detailed in the forthcoming sections. See \cite{Naiman17} for a simple breakdown of the Houdini graphics user interface (GUI) and a typical workflow session in an astrophysical context, as well as more general background and usage of Houdini. \par

Houdini accepts data formats which includes, but is not limited to, .geo, .bgeo, .json, .pdb, .obj, and .vdb. Once the simulation dataset is imported from a local directory and into Houdini via a File node in the Network View panel (as referenced via its path-to-file in the ``Geometry File'' parameter), one can examine all the attribute data---the different parameters included in the simulation proper such as temperature, density, position (x,y,z), etc.---via the ``Geometry Spreadsheet'' tab. This attribute data will later be used in the clustering algorithms. \par

\subsection{Clustering Algorithms}\label{subsec:algorithms}

Because this work is crucially dependent on choosing the appropriate clustering algorithm, the Gaussian Mixture Model (GMM)\footnote{\url{https://scikit-learn.org/stable/modules/mixture.html}} used in this work is explained thoroughly below. Other common clustering algorithms are explained in the appendix.


A popular and powerful unsupervised learning technique to cluster data is in the form of mixture models---probabilistic models for estimating in which subpopulation within an overall population a datum resides. In GMMs, subpopulations are construed to be Gaussian distributions with unknown parameters, such that all data (the ``population'') is thought to be generated from a finite mixture of these smaller distributions. Thus, for any one particular data point, there is an inherent probability to which subpopulation, or cluster, it belongs. In other words, a GMM is a parametric probability density function (PDF) for which its components are a sum of weighted Gaussian densities \citep{Reynolds09}. \par

For the algorithm to generate the requisite number of Gaussian mixtures, the user must first assign the number of clusters $M$, meaning that the number of clusters are known \textit{a priori} or assumed to be known. For GMM clusters to have an ascribed physical meaning, it has to be a reasonable assumption that the dataset can be generated from a superposition of Gaussian distributions. \par

The GMM mathematical formulation is described below, following the notation of \citep{Reynolds09}. The equation describing the probability of each point (in our case, a particle) being generated by each Gaussian component of the population $p(\textbf{x}|\lambda)$ can be written as 
\begin{equation}
    p(\textbf{x}|\lambda) = \sum_{i=1}^{M} w_i  g(\textbf{x}|\mu_i, \Sigma_i),
\end{equation}
where $\textbf{x}$ is some $D$-dimensional data vector particular to the dataset, $\lambda$ is the collection of variables parameterizing the model
\begin{equation}
    \lambda = (w_i, \mu_i, \Sigma_i), \hspace{0.5cm} i = 1...M,
\end{equation}
where $w_{i=1...M}$ are the individual weights of $M$  Gaussian components (constrained to sum to 1), and $g(\textbf{x}|\mu_{i}, \Sigma_{i})$ for $i$ = 1...M are the various Gaussian component densities. These densities are $D$-variate Gaussians formulated via
\begin{equation}
    g(\textbf{x}|\mu_i, \Sigma_i) = \frac{1}{(2\pi)^{D/2} |\Sigma_i|^{1/2}} \text{exp}\left( -\frac{1}{2} (\textbf{x}-\mu_i)^T \Sigma_i^{-1} (\textbf{x}-\mu_i) \right),
\end{equation}
where $\mu_i$ is the mean vector and $\Sigma$ is the covariance matrix. The covariance matrices can be of several types: `full', `tied', `diagonal', and `spherical'. In brief, `full' means full rank covariance, where each component has its own general covariance matrix; `tied' forces all components to share the same covariance matrix; `diag' allows for each component to contain their own diagonal covariance matrix; and `spherical' represents the case where there is a single variance for each component. A `full' rank covariance was used for this work. \par 

With the type of covariance selected, GMMs will then estimate the various parameters (the mixture weights, means, and covariances, all assumed within $\lambda$) using the Expectation-Maximization (EM) algorithm. EM is an iterative algorithm specifically designed to always converge to a local optimum, where parameter values of unobserved latent variables (in this case, the Gaussian components) are estimated by maximizing the likelihood \citep{Dempster77}. As the name suggests, there is an expectation and a maximization step. After random initialization of the parameters describing the components, the expectation step establishes a function representing the log-likelihood of the data based on those parameters, and by proxy, the latent distribution. In the case of GMM, a probability will be computed for each point being generated by each Gaussian component of the population, $p(\textbf{x}|\lambda)$. The intermediate goal of EM is to find some new model, $\lambda^*$, for which $p(\textbf{x}|\lambda^*)$ $\geq$ $p(\textbf{x}|\lambda)$. To achieve this, the maximization step will subsequently tweak the current estimate of the parameters to maximize the log-likelihood established from the expectation step. These same parameters will be used to form the new distribution of the unobserved latent variables (the Gaussian components) in preparation for the next E step. This process continues for a user-specified number of iterations (200 for this work) until some best guess solution is made. Each of the final Gaussian components whose parameters ultimately maximize the log-likelihood, are then defined as the clusters of the dataset. \par   

GMM is best used on flat geometries\footnote{An overview of clustering algorithms and general usecases can be found at: \url{https://scikit-learn.org/stable/modules/clustering.html}} (obeying Euclidean geometry when measuring distances), and will not inherently bias the cluster sizes to favor particular structures over others. However, it is important to ensure that each mixture has a sufficient number of datapoints (in tandem with choosing a reasonable number of components); otherwise, the covariance matrices become increasingly difficult to estimate, causing spurious estimates of infinite log-likelihoods. A metric to help determine the optimal number of components to describe the data is the Akaike information criterion (AIC) or the Bayesian information criterion (BIC). The particular number of components that produces the lowest AIC or BIC score is potentially the best option to use. \par

In this work, we chose a 5-cluster GMM with `full' covariance type, initialized by a random seed.

\section{Methods}\label{sec:Methods}

We now outline the procedure for pre-processing the simulation data, evaluating clustering results, and building a shader within Houdini informed by clustering results.

\subsection{Pre-processing}\label{subsec:Preprocessing}

Cinematic visualization can be computationally expensive. Here we impose some thresholds on the simulation data to save computational costs without devaluing the visualization. In this example, the thresholds were chosen based on \textit{a priori} knowledge of simplifying the data in the context of its visualization. This pre-processing is identical to that done by the AVL team in their ``Imagine the Moon'' dome show visualization, to allow for a side-by-side comparison. Their pre-processing criteria was, in part, motivated by the fact that a large fraction of the simulation volume had slowly-varying material on the outskirts, whose detailed behavior was not critical to understanding the evolution of the synestia's central regions. Thus, processing widely-spaced SPH sample points would have dominated the computation needed for rendering while adding little to the quality of the visualization. \par

In this work, we threshold two attributes from the simulation: smoothing length (a parameter used to control interactions between particles in SPH (see e.g. \cite{Springel01}), and density. At very high smoothing length values ($\gtrsim$ 100), the sphere sprites\footnote{Sprites, or 2D images set in a larger 3D scene, can be attached to particles such that the sprite image always faces the camera \citep{ORourke98}.} of the particles become too large, and subsequently dominate the visualization. However, such particles mostly reside on the outer fringes, and do not constitute the key components of interest. By thresholding to only have particles with $smoothLen$ $\textless$ 90, we saved on computational costs without compromising the visualization. \par

We also imposed an upper-limit density cutoff of $\textless$ 3.4 g/cm$^{3}$; because the densest particles are mostly within the metallic core of the post-impact body and so add no visual difference to the final render. Only particles that met the aforementioned smoothing length and density thresholds were used in this work, totalling to 35,987 from the original 100,989. \par

\begin{figure}
\centering
\includegraphics[width=\columnwidth]{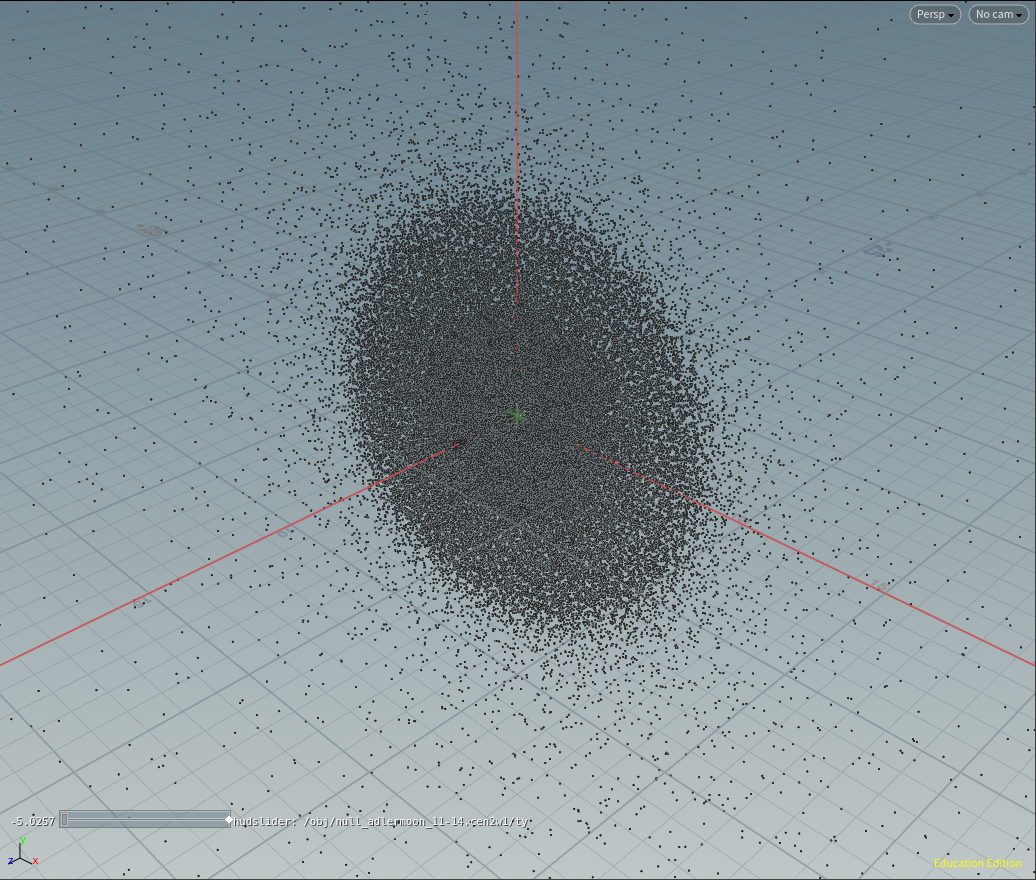}
\caption{The Scene View of the synestia, when data is first loaded into the Houdini software. Each individual grey particle is one of the 35,987 SPH particles after pre-processing (Section~\ref{subsec:Preprocessing}).}
\label{fig:synestia_viewport}
\end{figure}

The pre-processed dataset when loaded into Houdini appears as an agglomeration of particles, as shown in Figure~\ref{fig:synestia_viewport}.
%

\subsection{Importing clustering results into Houdini}

We tested several clustering algorithms in an attempt to find ``physically-interpretable'' clusters in the pre-processed data; that is, clusters that correspond to significant physical structures within the post-impact body, and not arbitrary, mathematical curiosities. This is important because non-significant or non-real structures emphasized in a final visualization can lead to false conclusions when interpreting the data. \par

We scaled our attribute values to be of the same order. This is necessary because some machine learning algorithms can be biased towards larger or smaller quantities, as some inherently assume that all attributes are centered about zero and have the same order variance. Thus, if a particular attribute has a variance several orders of magnitude larger than another, it might dominate the objective function and inhibit the estimator. Having performed many clustering tests in different phase-space permutations, we chose a temperature-specific entropy phase space because clusters were easily differentiated and could be physically interpreted. Note that temperature and entropy are conjugate thermodynamic variables and so provide a complete thermodynamic description making them well suited for describing the thermodynamic phase space. In the units of our example simulation, temperature ($\mathcal{O}(10^3)$ K) is four orders of magnitude smaller than the given specific entropy values ($\mathcal{O}(10^7)$ erg g$^{-1}$ K$^{-1}$). We scaled these values so that both variables are of the same order using the \texttt{StandardScaler} package from the \texttt{sklearn.preprocessing} module. Now any of the various clustering algorithms could be used. 

\begin{figure*}
\centering
\includegraphics[width=18cm]{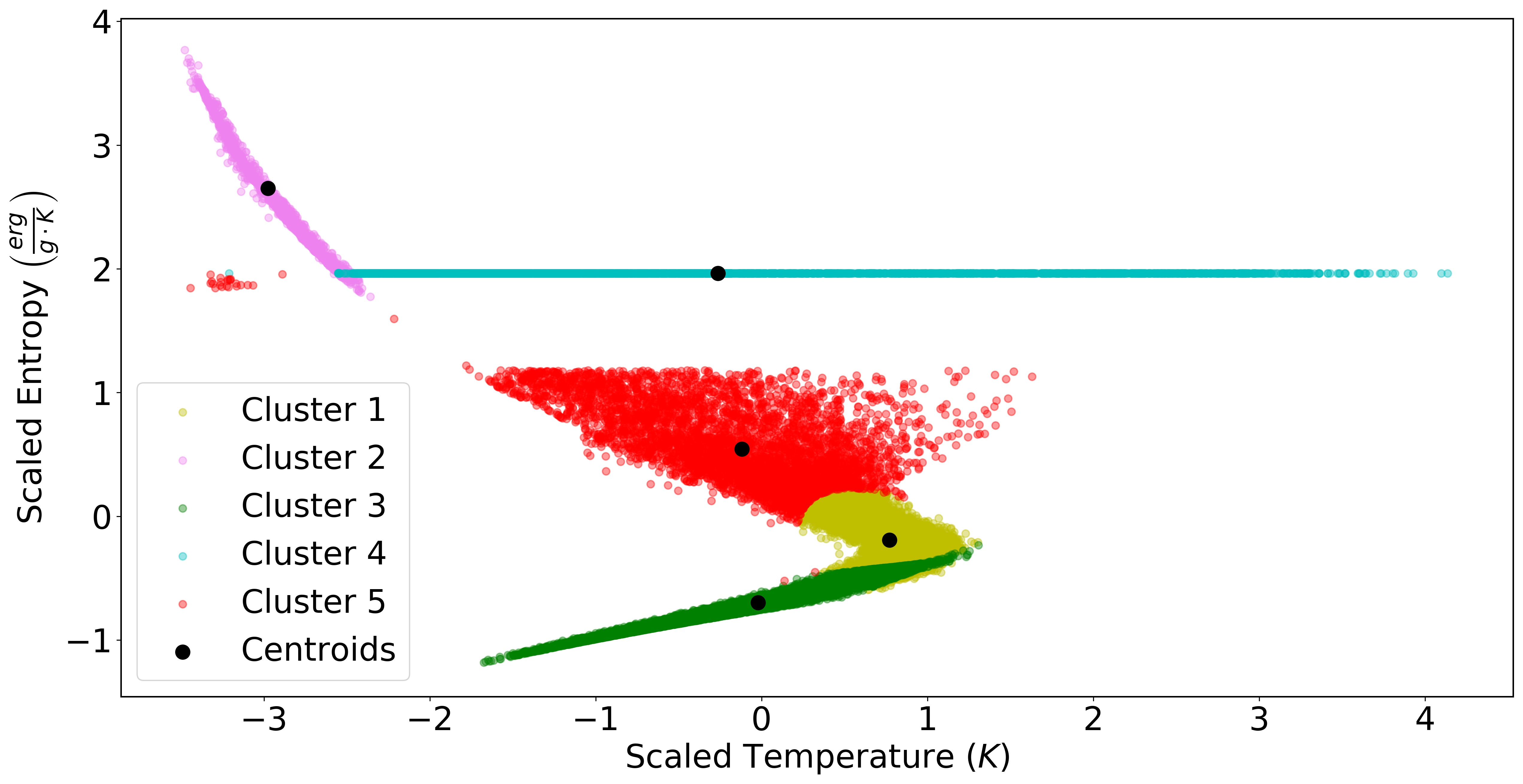}
\caption{The results of the 5-cluster GMM algorithm of the 35,987 SPH particles remaining after pre-processing, for a maximum of 200 iterations in a scaled temperature-specific entropy phase space. The covariance type was set to `full', with weights initialized by `kmeans' with a random seeding. Each cluster is represented by a different color with the centroids marked by black circles.}
\label{fig:5_clus_GMM_result}
\end{figure*}

Of all the clustering tests we performed, the 5-cluster GMM model was the ``best'' choice, in that each cluster was found to have a corresponding physically interpretable meaning. Moreover, a 5-cluster model is simple and sufficient to describe the data. What each cluster represents will be discussed in Section~\ref{sec:Results}. The 5-cluster GMM result is shown in Figure~\ref{fig:5_clus_GMM_result}. \par
\begin{figure*}
\centering
\includegraphics[width=18cm]{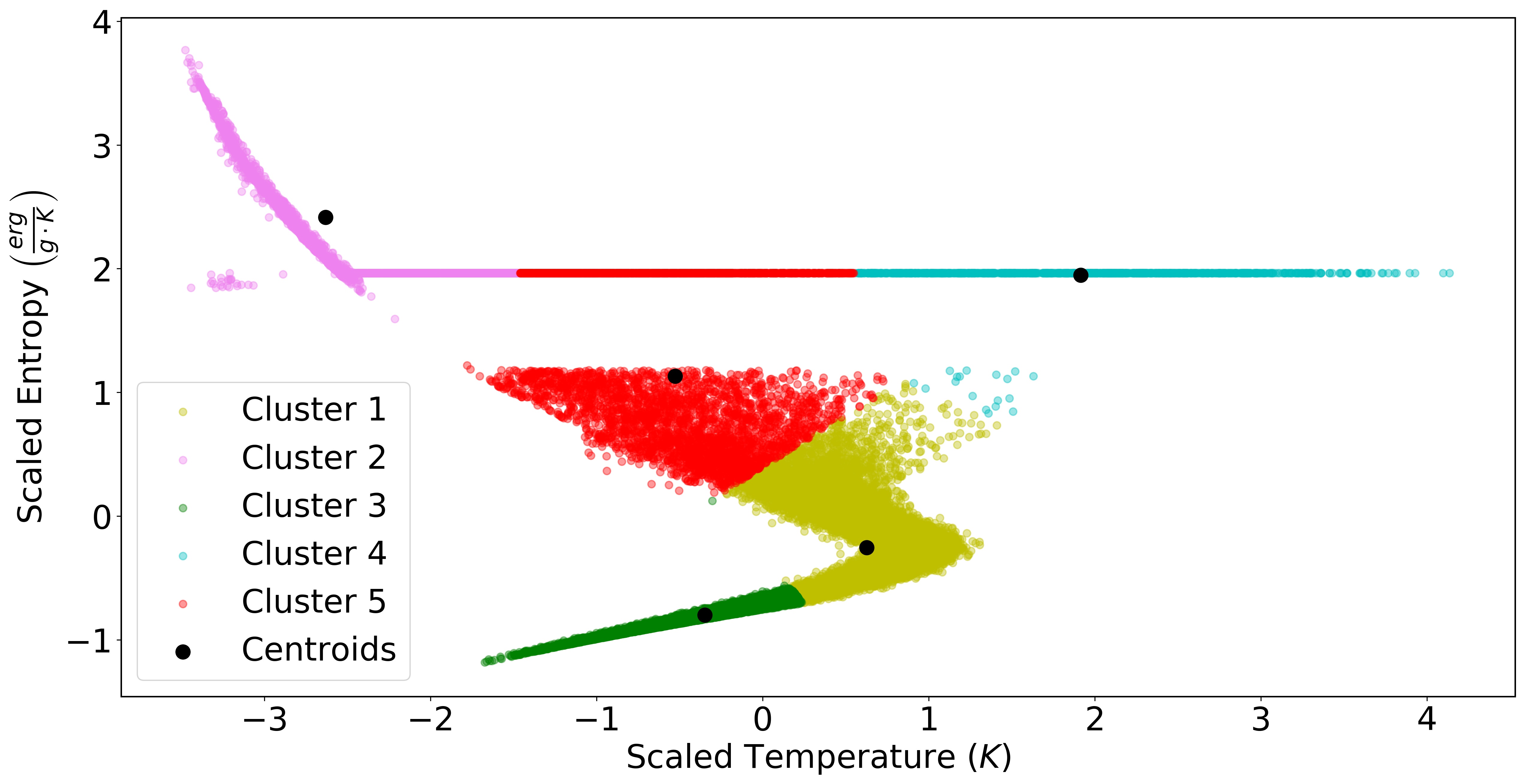}
\caption{The results of a 5-cluster Kmeans algorithm, for a maximum of 300 iterations in a scaled temperature-entropy phase space. The weights were initialized by `k-means++' in \texttt{sklearn.cluster.KMeans}, with 50 runs set by different centroid seeding. Each cluster is represented by a different color with cluster centroids marked by black circles. After pre-processing, there are 35,987 SPH particles to be clustered, ran with the same random seed as in the GMM clustering algorithm, the results of which are shown in Figure~\ref{fig:5_clus_GMM_result}. This is an example of a poor clustering algorithm choice in this phase-space because it arbitrarily splits likely physical clusters (such as a singular isentropic structure) into multiple disparate clusters. Further, Kmeans is designed to be used on flat geometry, but in this phase-space the data is non-flat.}
\label{fig:5_clus_Kmeans_result}
\end{figure*}

\begin{figure*}
\centering
\includegraphics[width=18cm]{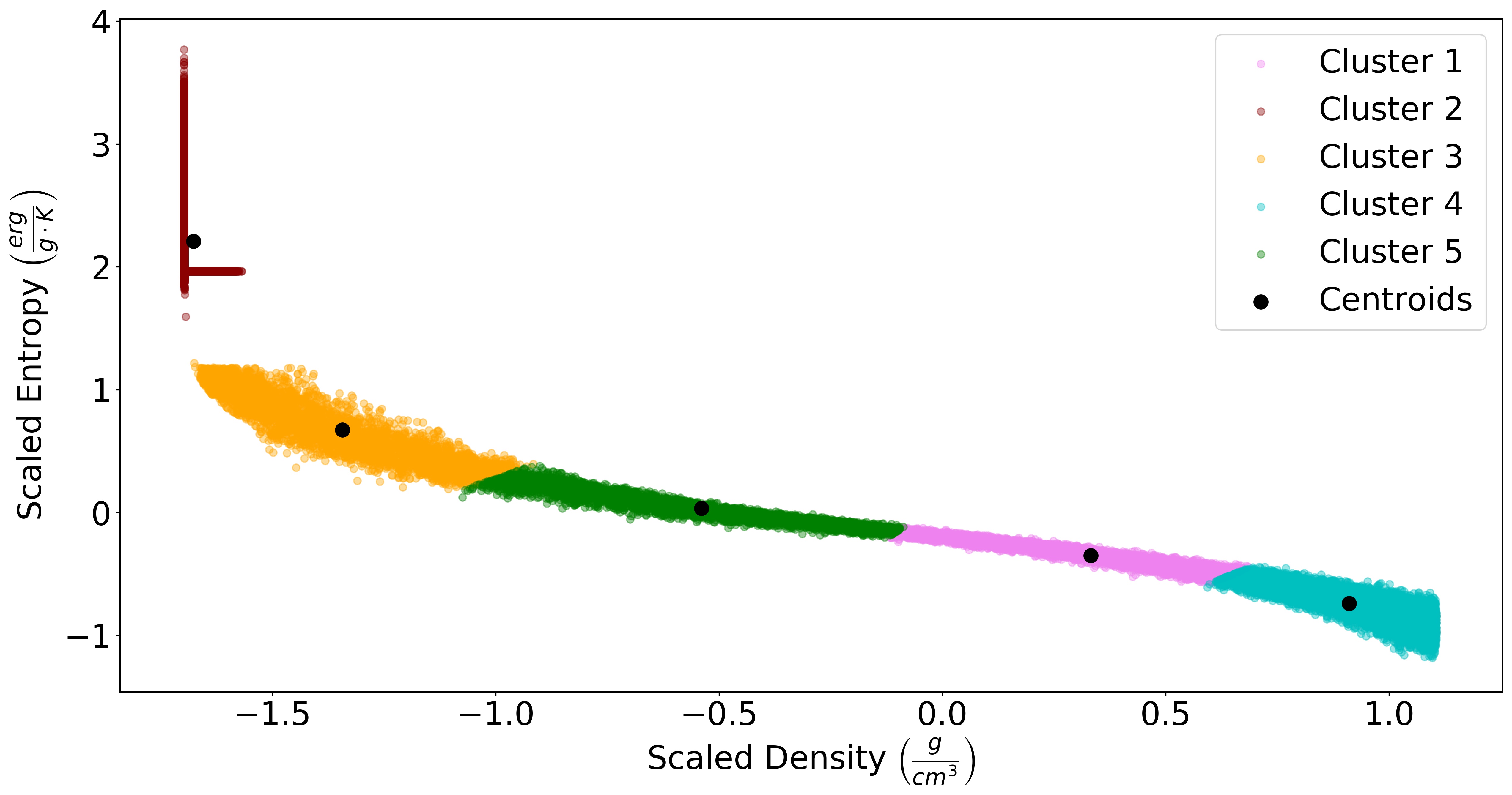}
\caption{As Figure~\ref{fig:5_clus_Kmeans_result}, but with the Kmeans clustering performed in a scaled density-entropy phase-space instead of a scaled temperature-entropy phase-space. This is also an example of a poor phase-space choice because most of the data is in one streamlined band, and it is difficult to determine where one cluster begins and another ends. Also, one cluster (red) incorrectly groups together two known significant structures: a vapor dome region (red vertical band), and an isentropic pure-vapor region (red horizontal band).}
\label{fig:5_clus_Kmeans_ent_rho_result}
\end{figure*}

Examples of poor clustering algorithm choices are shown in Figures~\ref{fig:5_clus_Kmeans_result} and~\ref{fig:5_clus_Kmeans_ent_rho_result}. Figure~\ref{fig:5_clus_Kmeans_result} shows a 5-cluster Kmeans algorithm in the same temperature-entropy phase space. In this case, data points are split seemingly arbitrarily. For instance, a single cluster of known significance---the outer material prescribed to be isotropic---is split into three clusters. This is likely due to a mis-application of the algorithm---Kmeans is designed to work for a flat geometry (flat manifold) usecase, whereas the data plotted in this temperature-entropy phase space is non-flat. \par
A second example of a poor choice for clustering was clustering in a density-entropy phase space. As seen in Figure~\ref{fig:5_clus_Kmeans_ent_rho_result}, the SPH particles in this phase-space are in a single ribbon-like stream. It is difficult to determine where one cluster (and hence an associated physical structure) ends and another begins with density smoothly decreasing with pressure. Further, this clustering attempt assigns one cluster (red) to what we know to be two different structures: an outer vapor dome region (vertical band) and an isentropic pure-vapor region (horizontal band). \par
Once a final clustering result was chosen (Figure~\ref{fig:5_clus_GMM_result}), the clustering ID results were imported into Houdini to inform the visualization. To do so, we wrote out the \texttt{predict} method values to a text file, and imported these values into Houdini via a ``Table Import'' node. Once these cluster ID values are assigned to an attribute value of the geometry, it becomes part of the dataset---all particles in the data are assigned a custom attribute value named ``map\_id\_to\_clus'', with an integer value ranging from 1-5, with the number representing their cluster ID. \par

\begin{figure*}
\centering
\includegraphics[height=10cm]{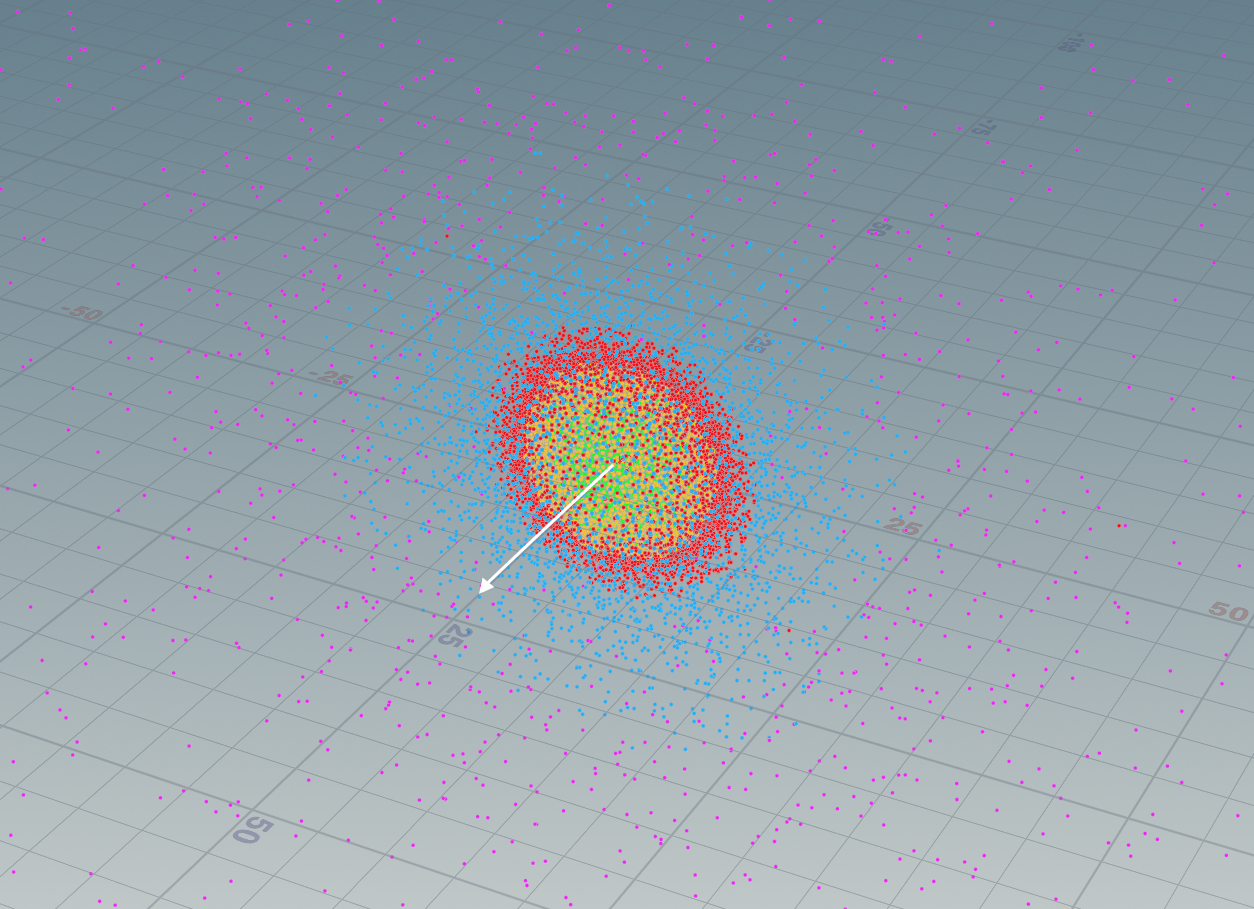}
\caption{Scene View of the synestia dataset, colored by its clustering ID results from the 5-cluster GMM result of Figure~\ref{fig:5_clus_GMM_result}. Thus, the color corresponding to each particle in 2D phase space can be seen in 3D position space in the Houdini software. The white arrow marks the rotation axis.}
\label{fig:Scene_View_GMM_5_clus}
\end{figure*}

\begin{figure}
\centering
\includegraphics[width=\columnwidth]{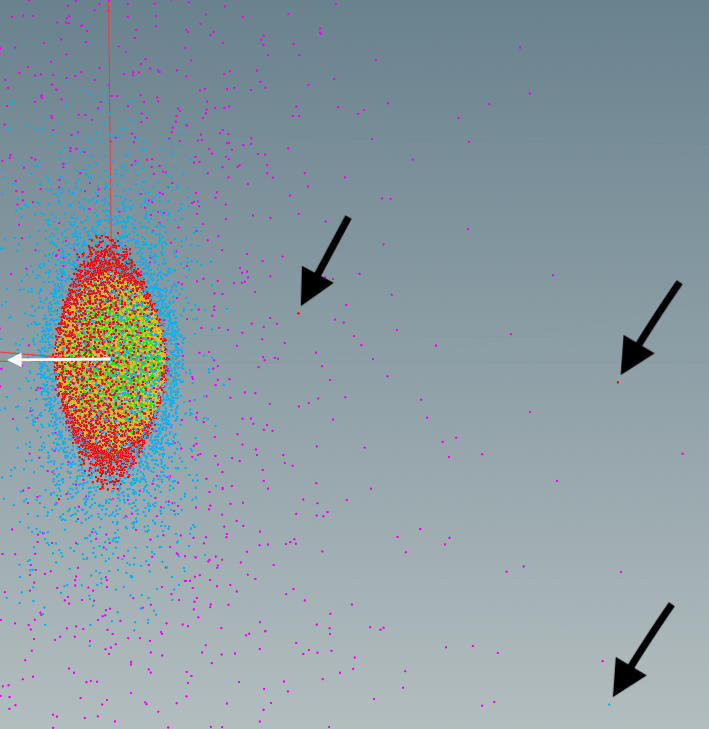}
\caption{A zoomed-in, rotated view from Figure~\ref{fig:Scene_View_GMM_5_clus}, where the rotation axis is along the horizontal extending out from the synestia bulge, as shown by a white arrow. On this scale, we can see individual particles and their cluster association. Individual outliers are seen, whereby an outlier is defined as being a particle with a different association relative to surrounding particles in physical space, marked by the large black arrows. These outliers are from the left-most grouping of red and cyan particles at (-3,2) in Figure~\ref{fig:5_clus_GMM_result}, where these few particles are possibly assigned an incorrect cluster.}
\label{fig:GMM_5_cluster_outlier}
\end{figure}

After a few more steps in Houdini (outlined in the Estra.ipynb notebook), we can display the particles shaded by their cluster assignment in the Scene View, as seen in Figure~\ref{fig:Scene_View_GMM_5_clus}.
This view shows the 3D spatial distribution of the clusters which had been identified in 2D phase space. This step is imperative to determine if the particular clustering algorithm used correctly identifies clusters representative of physical structures within the data. Moreover, the Scene View in Houdini is interactive, such that the user can easily change the viewing angle (see Figures~\ref{fig:Scene_View_GMM_5_clus},~\ref{fig:GMM_5_cluster_outlier}), zoom in/out, and select certain groups of particles to fully understand the data for all time steps (if time-evolving), etc. Outliers in the cluster assignment can also be identified, and their potential impact, if any, on the final render can be determined. For example, three outliers are evident in Figure~\ref{fig:GMM_5_cluster_outlier}. 
The minimal outliers have no visible impact on the final render, so we do not consider them further. It is important to note that we cannot know the ground truth cluster assignments, so the usage of ``outlier'' simply refers to a different cluster assignment from the overwhelming majority of those around the particle in 3D position space. \par

Finally, with the clustering results imported into Houdini, we can build a shader and perform the visualization. \par

\subsection{Algorithmically generating colormaps from clustering results}\label{subsec:colormap}

An advantage of clustering the data is that it allows for the automatic creation of transfer functions or color ramps which are informed by physical structures in the data. Houdini uses the transfer function (a set of $(R,G,B, \alpha)$ values for a range of data values for a variable, which here is temperature) in shading the final image\footnote{While the standard definition of a ``transfer function'' deals with $(R,G,B, \alpha)$, Houdini splits this into a ``color ramp'' of $(R,G,B)$ values and a ``spline ramp'' for $\alpha$.}. In building a color ramp, the shader requires `position' markers called key points with associated $(R,G,B)$ values. For instance, although a file with color map data may have 1024 rows of $(R,G,B, \alpha)$ values for an emissive blackbody color scheme, Houdini requires only a handful of position markers containing $(R,G,B)$ values, and uses a user-defined interpolation function (e.g. Linear, B-Spline, Bezier, Catmull) to create the ramp. Then, once a range of temperatures are defined for that color ramp, each temperature in that range will be assigned a unique $(R,G,B)$ value in the shader. \par

For this work, each position marker with its unique $(R,G,B)$ value is informed by the clustering results. First, the entire temperature range is rescaled back to physical units (Kelvin) using the \texttt{inverse\_transform} method in \texttt{sklearn}. The range of temperatures is then normalized to be between 0-1 (as position markers in Houdini have values [0,1]). The minimum, mean (represented by the centroid), and maximum temperature values across all members of each cluster are assigned a `position' marker value from [0,1] based on where they lie on the normalized scale. Lastly, with each key points' position value determined, its corresponding $(R,G,B,\alpha)$ value in the imported color scheme is set\footnote{Where in this work a temperature range [0, 14308 K] was normalized to [0,1], a temperature of 6000 K would have a marker position value of 0.419. This would correspond to the 429th row of the 1024 row color map.}. In other words, position markers from each cluster correspond to Houdini key points on the color ramp. \par

We used a modified emissive color scheme from the AVL's catalog for our final renders. It maps any temperature value to its associated $(R,G,B, \alpha)$ color value using an emissive blackbody color temperature scale, and is non-perceptually uniform\footnote{If desired, there are other perceptually uniform color maps available to use from e.g. \cite{Moreland16} and \url{https://github.com/liamedeiros/ehtplot}. Some additional perceptually uniform renders of this dataset are found in the appendix.}. \par

\begin{figure*}
 \begin{minipage}[c]{1.0\linewidth}
  \centering
  \begin{center}
  \textit{(a)}\includegraphics[height=3cm, width=17cm]{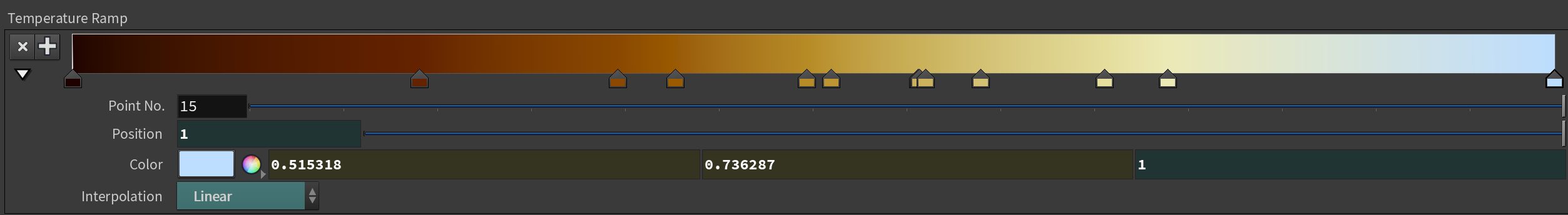}
  \textit{(b)}\includegraphics[height=3cm, width=17cm]{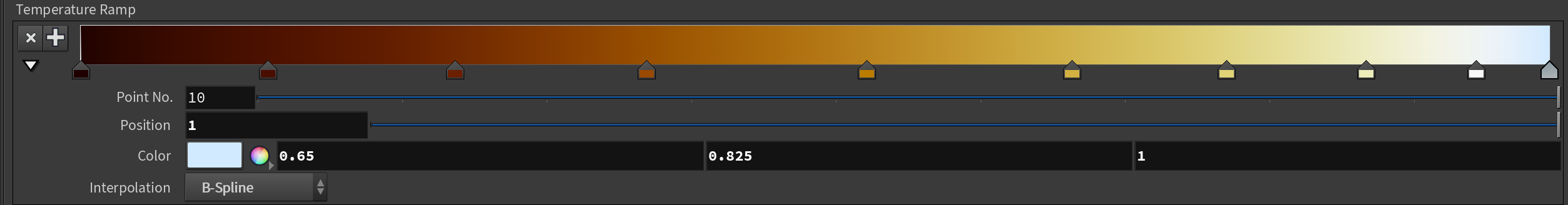}
  \caption{\textit{(a)}: The non-uniform, emissive colormap with markers and associated $(R,G,B)$ values informed by the 5-cluster GMM result. The colormap is over temperatures [0 K, 14308 K] to match that of the AVL rendering for later comparison, despite the dataset temperatures ranging from [2453 K, 14308 K] (the range [0 K, 14308 K] was chosen for aesthetic reasons). Here, the hottest temperatures map to light blue, moderate temperatures map to orange, and the coolest temperatures map to dark red/brown. \textit{(b)}: The custom AVL colormap, from [0 K, 14308 K], used in the ``Imagine the Moon'' dome show. Ten key points are used with a B-Spline interpolation, and are roughly equally spaced over the length of the colormap. Here the hottest temperatures map to mostly light yellow/white tones instead of a light blue like in the \texttt{Estra}-generated colormap.}
  \label{fig:temperature_ramp}
  \end{center}
  \end{minipage}
\end{figure*}

Our final clustering-informed temperature ramp is in Figure~\ref{fig:temperature_ramp}(a), and the AVL's custom temperature ramp (used in the ``Imagine the Moon'' dome show) is in Figure~\ref{fig:temperature_ramp}(b). \par
%

\subsection{Building a custom shader}

To create the visualization, Houdini requires a shader to tell its native renderer, Mantra, how light interacts with each particle in the scene.\footnote{Many other popular renderers such as Renderman will work, but may involve different steps to create a shader.} Once established, the scene can be rendered, where the three-dimensional scene is transformed into a two-dimensional image. Although there are built-in shaders, none of them are tailored to emissive, SPH data. In the ``From\_Results\_to\_Render.ipynb'' notebook, we detailed how we created a general-use particle sprite shader for emissive material, with values tailored to our example dataset. Future projects can simply copy this shader network into their own Houdini scene files, and modify the values appropriately. We summarize the process of building our shader below. \par

Because the material in SPH simulations is, as the name implies, divided into particles, it is best to design the visualization using particles. Thus, we transformed each data particle into a sphere sprite with a non-uniform rational B-spline (NURBS) primitive type\footnote{\url{https://www.sidefx.com/docs/houdini/model/primitives.html\#nurbs}}. Also, because each simulated SPH particle has a unique size or radius, we forced its sphere sprite to retain its smoothing length value from the data by assigning it to each particles' Houdini ``pscale'' value\footnote{While ``smoothing length'' is a data variable, ``pscale'' refers to a multiplier in Houdini on the size of a particle. We assigned the value of smoothing length to a particle's pscale. Conceptually these terms are different, but as far as implementation in Houdini is concerned, they are the same. We used the smoothing length to drive the size of the particle, by assigning it to pscale, or ``particle size multiplication factor''.}. Lastly we duplicated our pre-processing step as described in Section~\ref{sec:Methods} to ensure only the data retained in the clustering steps (and aligned in the same order) is used in the final visualization. \par

To build a custom shader, a Material Shader Builder node\footnote{\url{https://www.sidefx.com/docs/houdini/nodes/vop/materialbuilder.html}} was modified. Parameter nodes\footnote{\url{https://www.sidefx.com/docs/houdini/nodes/vop/parameter.html}} and ramps\footnote{\url{https://www.sidefx.com/docs/houdini/nodes/vop/rampparm.html}} were created to read in the data's temperature, density, and smoothing length values, because these control the emission color and opacity over a range of values. For example, a spline ramp type was used to adjust the transparency of each particle based on its smoothing length value. In this simulation, the largest particles (largest pscale) reside on the outer fringes of the simulation, as pscale is related to how many and how close its nearest neighbors are; the smallest particles/lowest pscale values constitute the central region of the post-impact body. Mapping large pscale values to low opacity/high transparency and small pscale values to high opacity/low transparency allows for the whole simulation to be seen and visualized; otherwise, the largest particles would dominate the visualization, and obscure the rest of the data. \par

\begin{figure*}
\centering
\includegraphics[height=3cm, width=17.5cm]{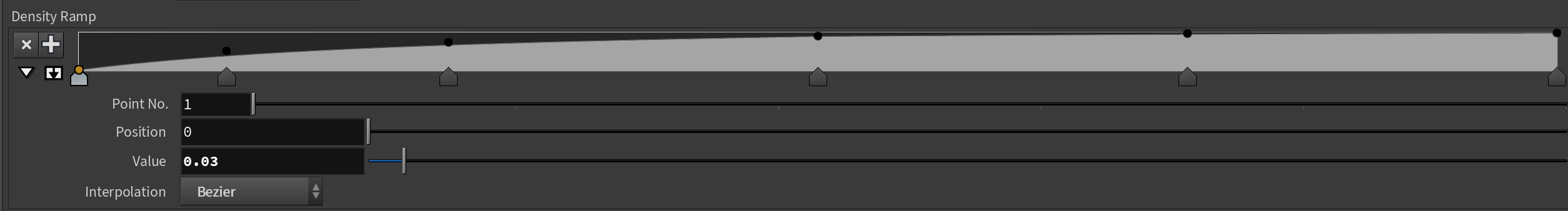}
\caption{The density ramp, mapped to opacity in the shader, where density values increase from left to right from 0 to 3.4---its max value after pre-processing. The higher the `Value' indicated by the ramp, the more opaque (less transparent) each particle at that particular density will be. Values are determined by a cubic Bezier curve with the control points $\textbf{P}_0 = (0,0)$, $\textbf{P}_1 = (0,0.2)$, $\textbf{P}_2 = (0,1)$, and $\textbf{P}_3 = (1,1)$. The `Interpolation' between each point is also set to `Bezier'. Less dense particles (typically near the outer edges of the synestia) are less opaque, and more dense particles (typically concentrated in the center of the synestia) are more opaque.}
\label{fig:density_ramp_bezier_curve}
\end{figure*}

In choosing the mapping transfer function for the temperature, density, and smoothing length opacity ramps, we made simple assumptions to enable ease of use. We assumed the density (opacity) ramp to have the form of a cubic Bezier curve, governed by the parametric equation:
\begin{equation}
\begin{split}
    \textbf{B}(t)=(1-t)^3\textbf{P}_0 & + 3(1-t)^2 t \textbf{P}_1 + 3(1-t) t^2 \textbf{P}_2 \\  
    & +t^3 \textbf{P}_3, 0 \leq t \leq 1,
    \label{eq:CubicBezierCurve}
\end{split}
\end{equation}
with control points $\textbf{P}_0 = (0,0)$, $\textbf{P}_1 = (0,0.2)$, $\textbf{P}_2 = (0,1)$, and $\textbf{P}_3 = (1,1)$, where $t$ is time (to trace out the full curve, not to be confused with a physical time associated with our dataset). These control points define a control polygon from which our curve is drawn. This resulting curve represents values for the position and value markers on the density ramp. We chose five values to define the ramp traced out by the curve: (0, 0), (0.1, 0.53), (0.25, 0.75), (0.5, 0.91), (1,1), as shown in Figure~\ref{fig:density_ramp_bezier_curve}\footnote{The ramp starts at an arbitrary non-zero y-value (0,0.03) so that the particle(s) with the lowest densities still contain some opacity, and are not completely transparent in the render.}.
These specific control points were chosen because: 1) the rectangular box of the ramps have domains $x \in [0,1]$, $y \in [0,1]$; 2) this choice results in a good balance of opacity as a function of density, whereby particles of all densities are visible, and are not obscured due to the extreme values---the densest particles (in the central region of the body) are visible without drowning out the least dense particles (on the edges) and vice versa. \par 

\begin{figure*}
\centering
\includegraphics[width=16cm]{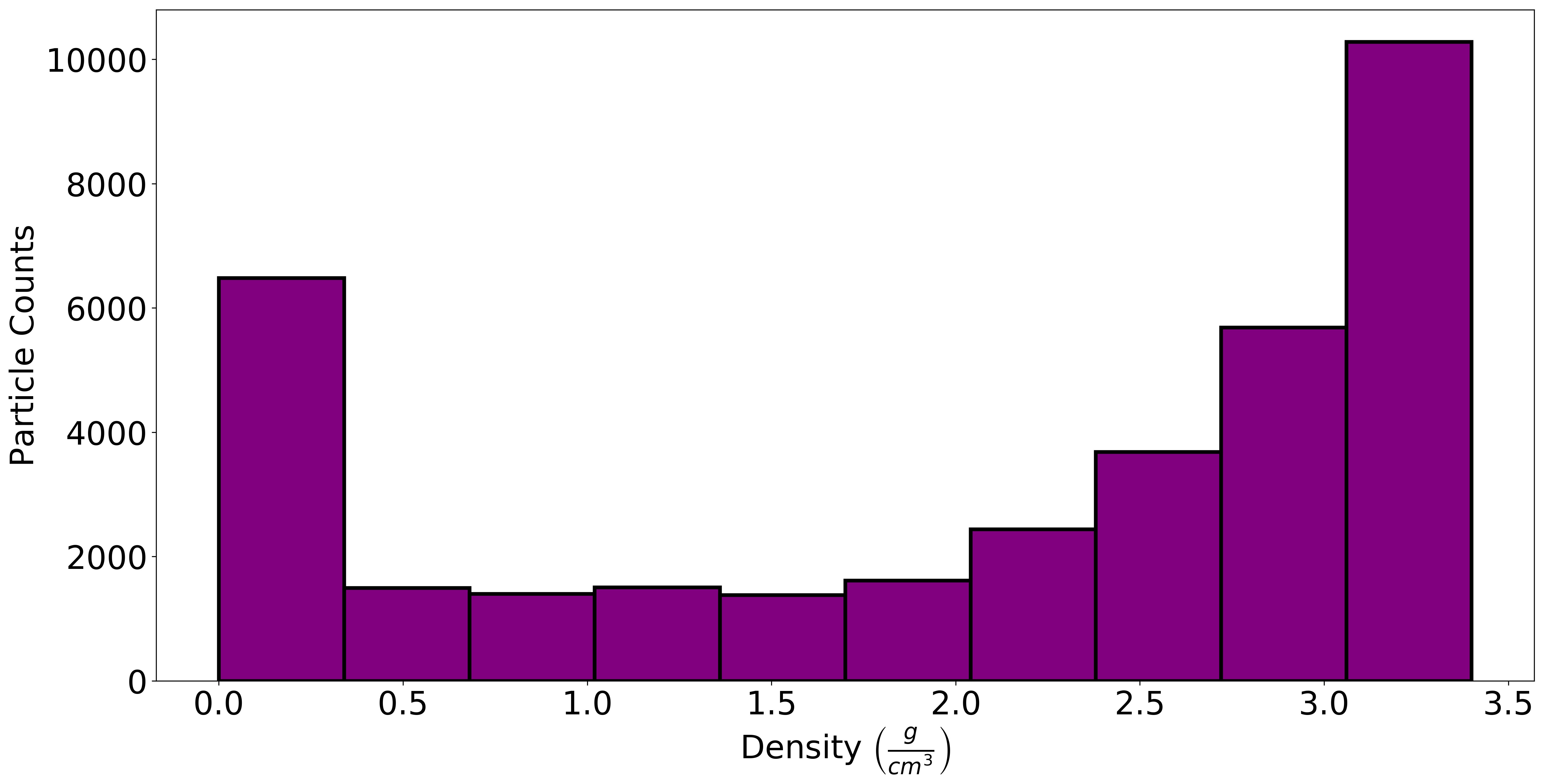}
\caption{A histogram of the number of particle counts as a function of binned (pre-processed) density. Most particles are either in the highest density regimes (center of post-impact body) or the lowest density regimes (out in the outer fringes). For our purposes, the higher density regions of the simulation are most interesting to visualize, in part to better compare this work's renders to those from the AVL.}
\label{fig:rho_histogram}
\end{figure*}

When plotting a histogram of the number of particle counts as a function of binned density, there is a double peak at the lowest and largest density values (Figure~\ref{fig:rho_histogram}).
A cubic Bezier curve is thus a good starting point to allow a particle to be simultaneously more opaque and emissive with increasing density (as this density ramp is connected to both opacity and emission multipliers), while allowing for the many contained lowest density particles in the center to be visible and not overpower the visualization. This cubic Bezier curve will be useful in many applications where dense objects are embedded in less-dense mediums, e.g. star formation. \par

The temperature ramp controls the color as a function of temperature. Its values are the result of the GMM clustering algorithm, which is the transfer function shown in Figure~\ref{fig:temperature_ramp}(a). \par

Lastly, the pscale ramp also controls opacity. Its values are the transposed values of the cubic Bezier curve used in the density ramp. This way, the largest pscale values (largest key position marker values on the ramp) have low opacity/high transparency (low ramp y-values), and vice-versa. \par

\begin{figure*}
 \begin{minipage}[c]{1.0\linewidth}
  \centering
  \begin{center}
  \textit{(a)}\includegraphics[width=0.30\textwidth]{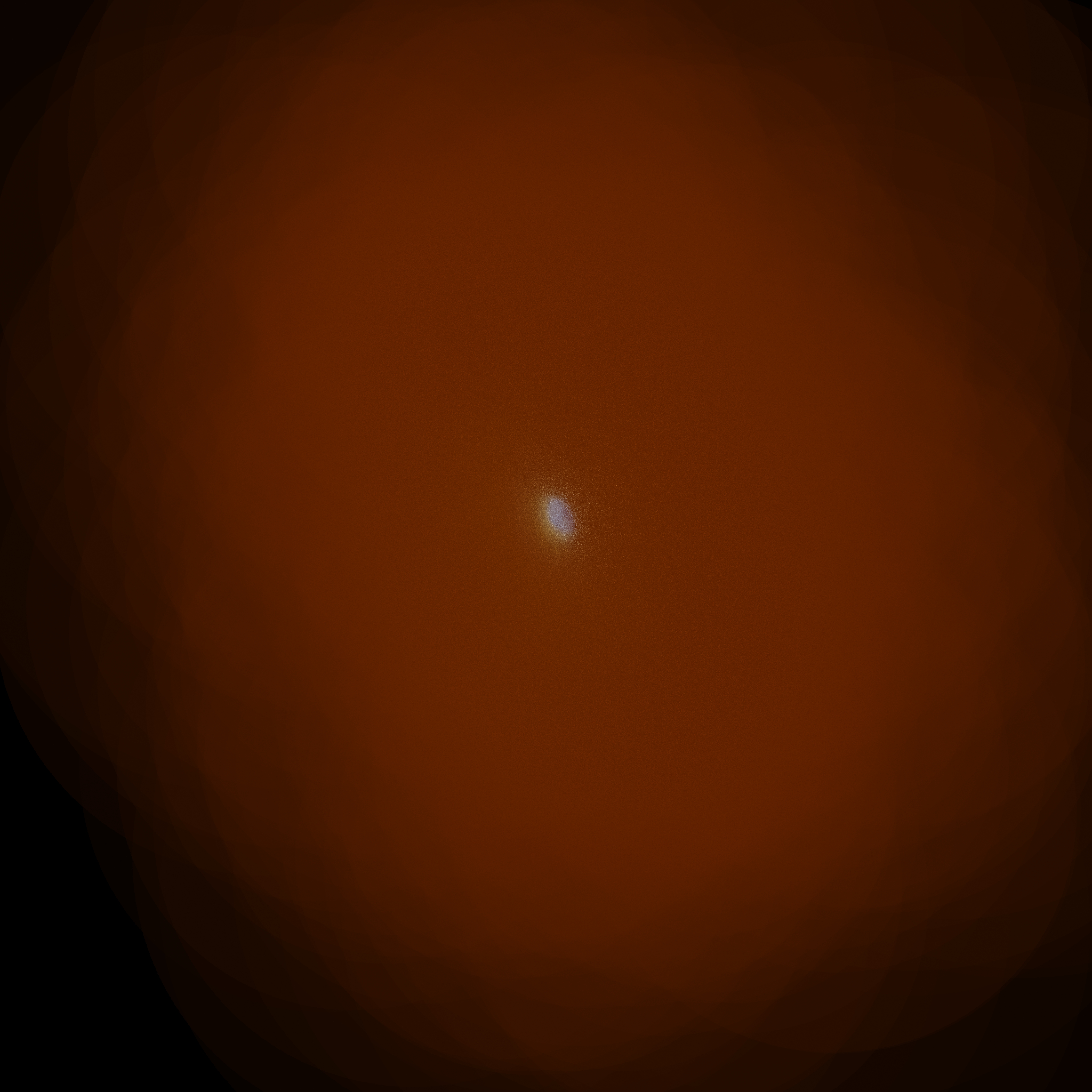}
  \textit{(b)}\includegraphics[width=0.30\textwidth]{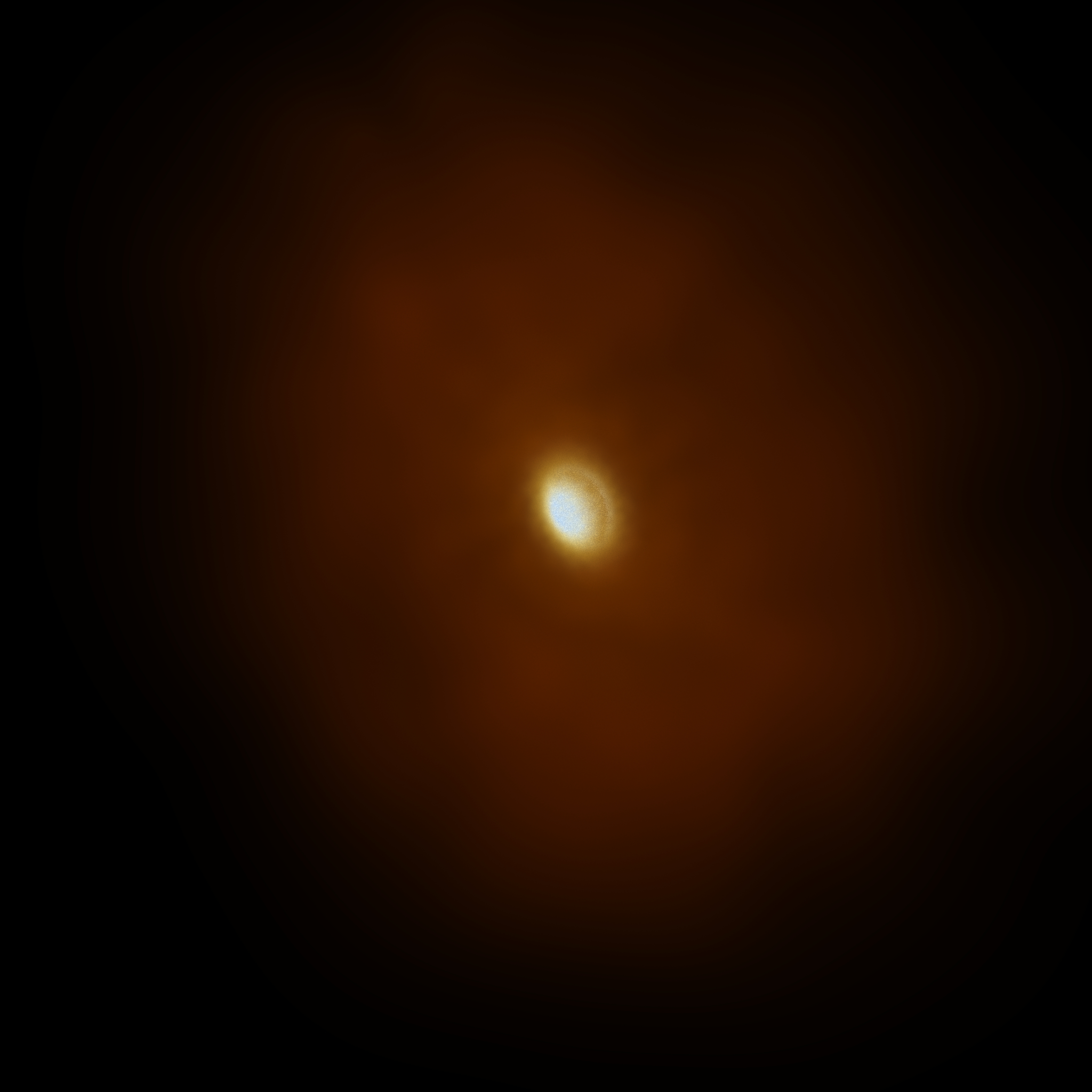}
  \textit{(c)}\includegraphics[width=0.30\textwidth]{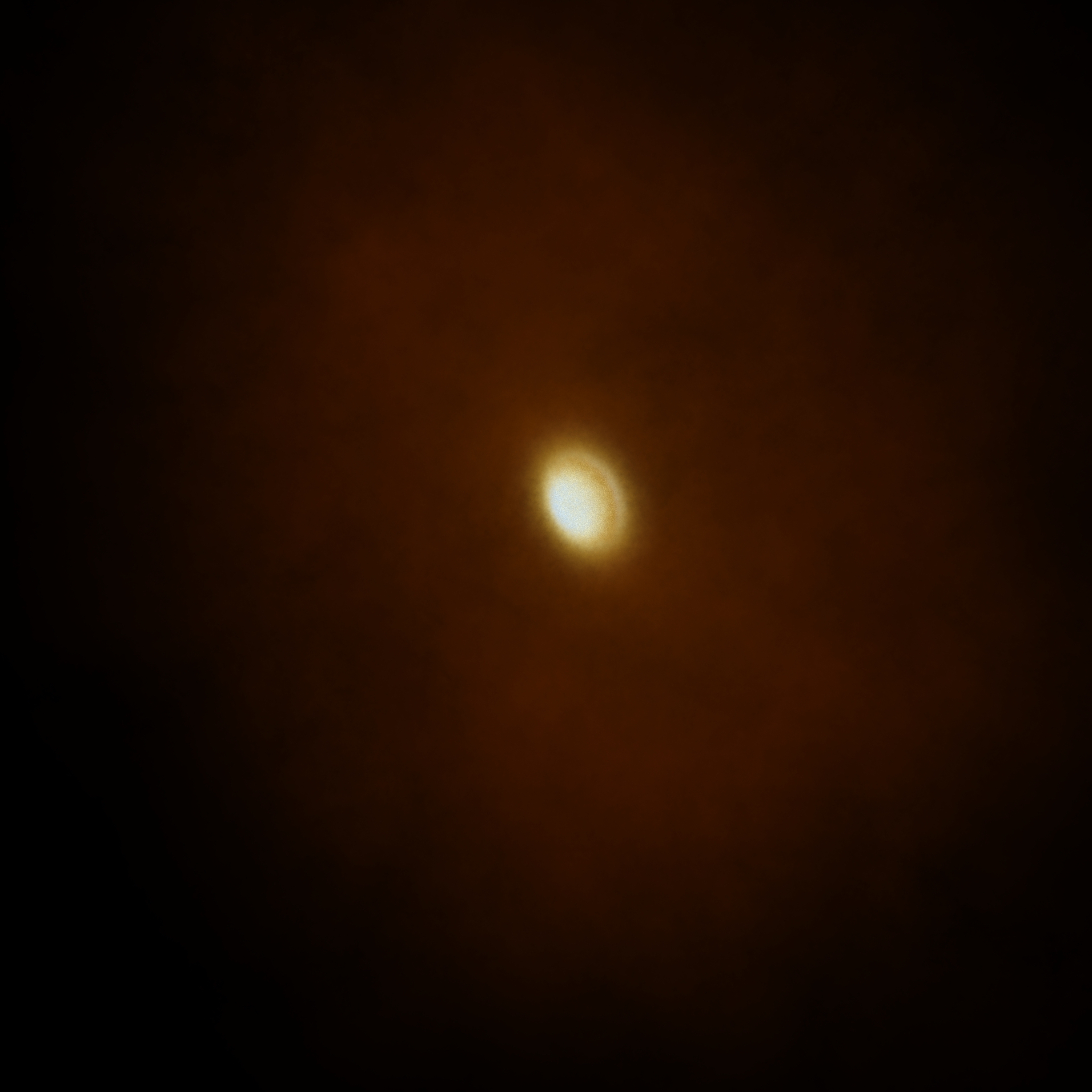}
  \caption{\textit{(a)}: This test render uses an incomplete shader, after temperature and pscale ramps are created. Visually it is evident from the color mapping that we have a redder, cooler outer region and a hotter, whiter/bluer inner region. \textit{(b)}: This test render, of the same scale and aspect ratio as \textit{(a)}, also uses an incomplete shader, but is after a screen space distance $d_{screen}$ is calculated and used in a Gaussian falloff profile for each sphere sprite. Because there is little clumping of the dust and gas clouds, randomized noise needs to added to visually suggest sub-grid scale fluctuations. \textit{(c)}: The final render, after random noise is added to complete the \texttt{Estra} shader. Because this view is distant from our areas of interest, a close-up of this same render is found in Figure~\ref{fig:Estra_AVL_render}(a).}
  \label{fig:progression}
  \end{center}
  \end{minipage}
\end{figure*}

Once the temperature and pscale ramps were set, a test image was rendered, shown in Figure~\ref{fig:progression}(a). This render has a hot inner region and a cooler outer region, but the look is clumpy and the center of the post-impact body is dim. With many of the interesting features occluded, more work was done.
%
%

Because luminance from each sphere sprite is isotropic and its intensity follows Lambert's cosine law, the screen-space distance from the viewing center was calculated to inform the intensity falloff. This falloff is dominated by the Gaussian function and is known as a Gaussian falloff profile. The screen-space distance $d_{screen}$ is $\sqrt{1-(\textbf{N} \cdot \textbf{I})^2}$, where $\textbf{N}$ is the (normalized) normal vector from the sprite surface, and $\textbf{I}$ is (normalized) vector for the incident light. In this manner, the screen-space distance is largest when $\textbf{N}$ and $\textbf{I}$ are orthogonal, i.e. in the line-of-sight of the viewing angle, and is smallest when $\textbf{N}$ and $\textbf{I}$ are unidirectional. A node network was made to perform the screen-space distance from viewing center calculation. Once completed, its output $d_{screen}$ value was used as an input in another node network to calculate a Gaussian falloff profile for each sphere sprite. This Gaussian falloff node network use the standard Gaussian function $g(x)$ of the form
\begin{equation}
    g(x) = a e^\frac{-(b-x)^2}{2c^2},
\end{equation}
where the parameter $a$ is the height of the curve's peak, $b$ is the position of the center of the peak (the $d_{screen}$ value) and $c$ (the standard deviation) controls the width of the ``bell''. A Gaussian profile is a good approximation for the spectral intensity profile, because each sphere is a resolved point source. The point spread function (PSF) that represents the light distribution of such a point source is approximately Gaussian \citep{Sterken92}. \par

With these new node networks added to our shader, another test render was created, and is shown in Figure~\ref{fig:progression}(b).
%
%


Finally, the clouds of gas and dust are randomized with noise to visually suggest sub-grid scale fluctuations, as a result of turbulent motion. These noise values, as well as other parameters such as the Gaussian profile constants, can be fine-tuned to achieve the purpose of the visualization. \par

With final noise values and Gaussian profile constants decided, our shader is completed and a final render is made, as shown in Figure~\ref{fig:progression}(c).
%
%

\section{Results \& Discussion}\label{sec:Results}


\subsection{Comparing Estra and AVL renders}

\begin{figure*}
 \begin{minipage}[c]{1.0\linewidth}
  \centering
  \begin{center}
  \textit{(a)}{\includegraphics[width=0.47\textwidth]{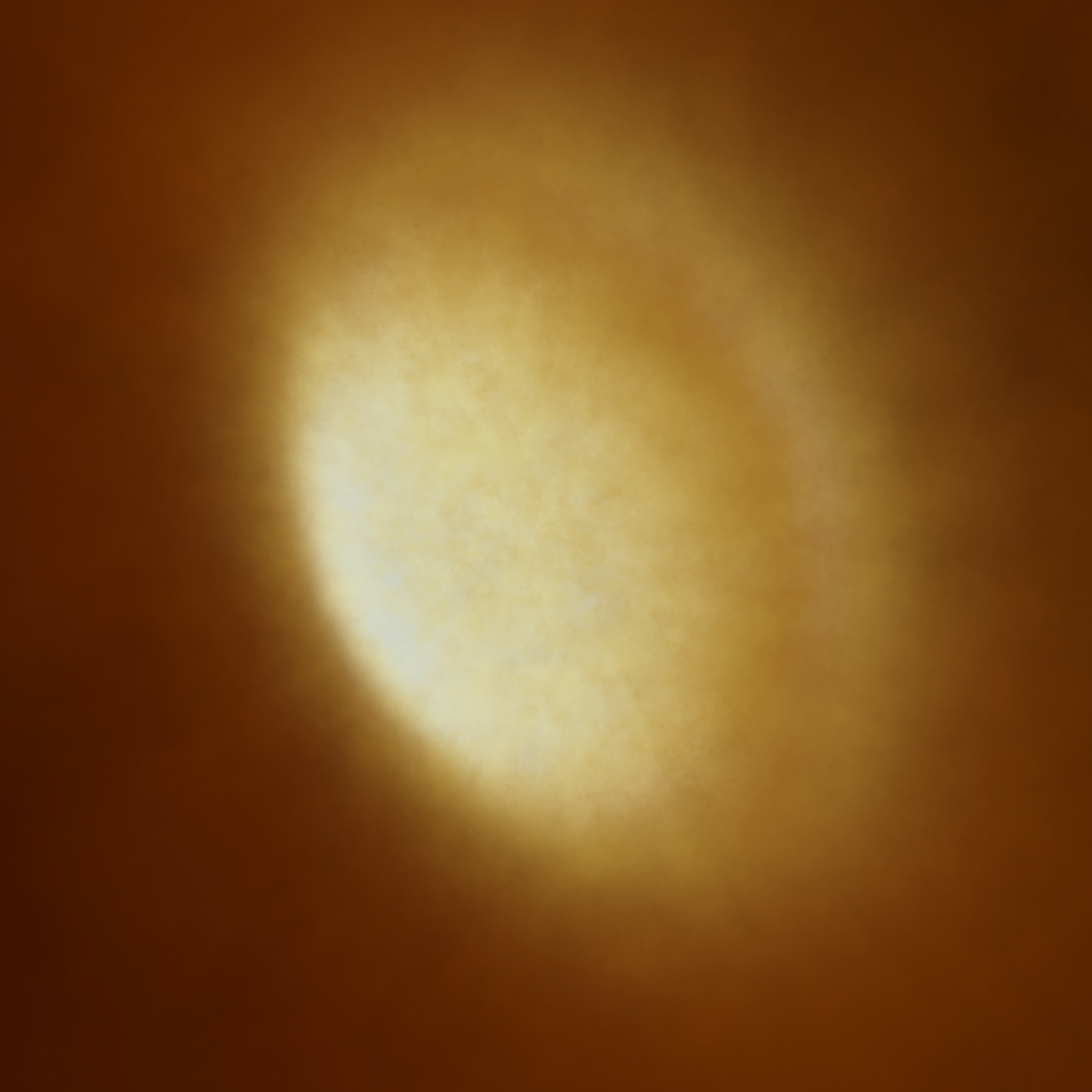}}
  \textit{(b)}{\includegraphics[width=0.47\textwidth]{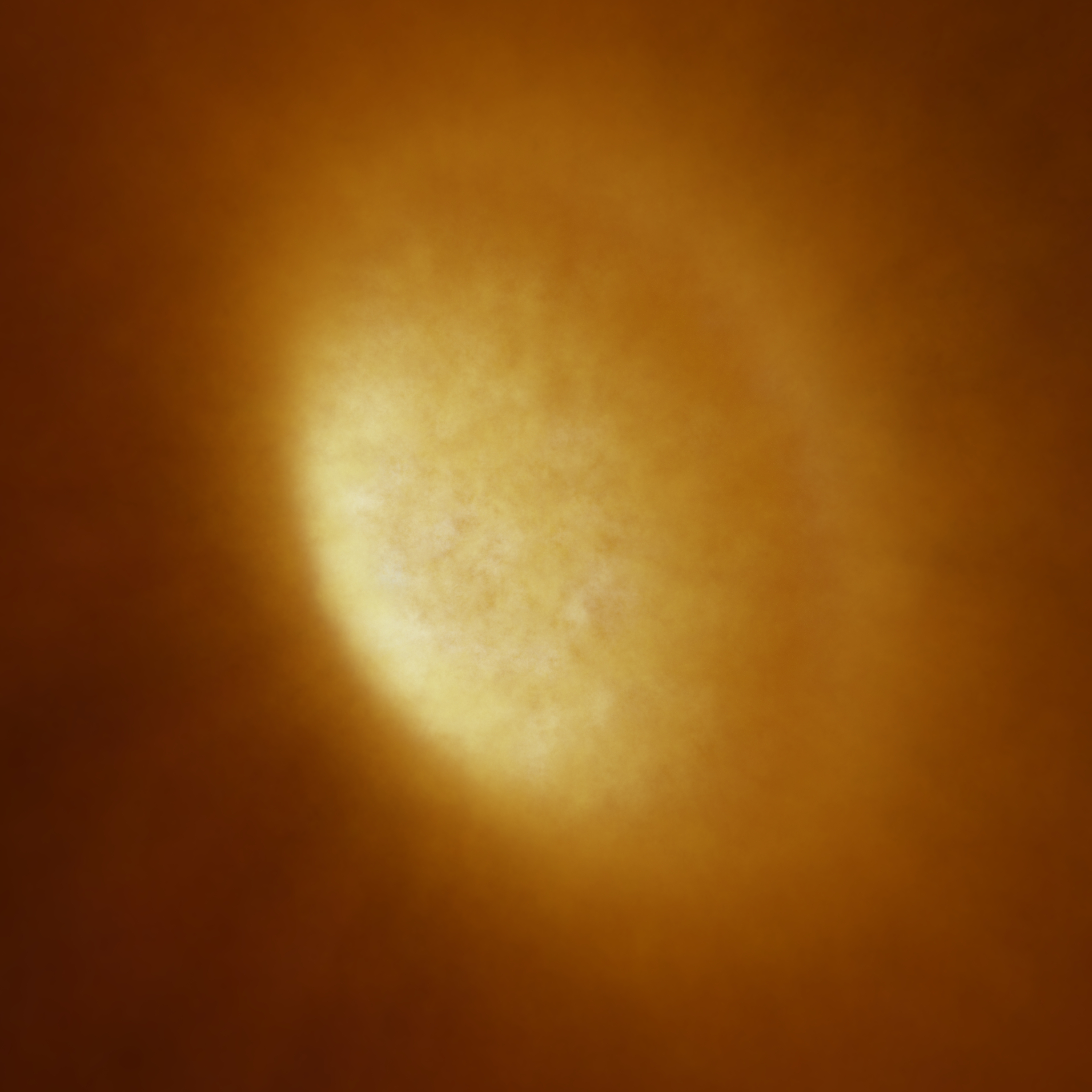}}
  \caption{The final renders of the synestia (of the exact view shown in Figure~\ref{fig:synestia_viewport}) using \textit{(a)} the \texttt{Estra} shader---with its colormap informed by the 5-cluster GMM result (Figure~\ref{fig:temperature_ramp}(a))---and \textit{(b)} the AVL's more complicated, custom shader not informed by a machine learning algorithm (colormap from Figure~\ref{fig:temperature_ramp}(b)). Although choosing the appropriate clustering algorithm and developing the pipeline took one of us (P. D. Aleo) approximately two months, $(a)$ was created in approximately a day's work once the workflow was established. Meanwhile, the timeframe for $(b)$ was similar, but involved the work of three visualization designers (A.J. Christensen, K. Borkiewicz, R. Patterson). This demonstrates how a scientist can create a simplified yet accurate visualization of their own work, comparable to the quality of one produced by a professional
visualization team.}
  \label{fig:Estra_AVL_render}
  \end{center}
  \end{minipage}
\end{figure*}

A zoomed-in view of Figure~\ref{fig:progression}(c), the final \texttt{Estra} render using GMM clustering which focuses solely on the center of the synestia structure, is shown in Figure~\ref{fig:Estra_AVL_render}(a). For comparison, the manually-designed AVL render, which utilizes a more complicated shader and is used in the full dome show ``Imagine the Moon'', appears in Figure~\ref{fig:Estra_AVL_render}(b) with the same scale and aspect ratio.
The $\texttt{Estra}$ rendering is qualitatively similar to the AVL rendering. Both retain the same bright bulge, and have a dusty ring of material on a plane perpendicular to the rotation axis. Additionally, both are emissive, and have the same clumping of gas and dust, with Gaussian falloff, and a similar color palette. The AVL rendering has more red tones, and less bright white highlights of the hottest material in the bulge, due to the slight difference in color maps. Although the color map used for The \texttt{Estra} rendering (Figure~\ref{fig:Estra_AVL_render}(a)) is a simplified and extended version of The AVL rendering (Figure~\ref{fig:Estra_AVL_render}(b)), it contains a different transfer function based on the clustering results. For example, moving the position markers (which themselves are directly determined from the minimum, mean, and maximum values of each cluster---see Section~\ref{subsec:colormap}), changes how temperature is mapped to a particular color. The spacing width between the position markers on the ramp controls the rate at which the color changes; the larger the spacing between the markers, the greater the change in color space and vice-versa. Hence, dissimilar phase-space clusters have a wider color palette spread for easier visual identification in the \texttt{Estra} render. \par

It is important to keep in mind that both renders cannot produce a fully realistic image; there is not one particular ``right'' or ``correct'' render. In fact, the purpose of this work is to enable a scientist with little to no background in cinematic visualization to visualize their own work, and we have shown that with clustering-informed color-mapping and simple assumptions, one can create compelling visualizations aesthetically similar to ones produced by visualization teams for full dome productions. \par

The \texttt{Estra} pipeline is also potentially useful in the production pipelines of visualization teams. They can create clustering-informed visualizations to reduce time spent on data exploration, and simply tweak or complicate them as necessary to match their needs. \par

Once the shader network is setup and pre-processing is completed, it is trivial to try any number of clustering algorithms and import the cluster IDs into Houdini. Building the general purpose shader network from scratch takes approximately an hour, and running a clustering algorithm and importing clustering results takes half that. The only significant time bottleneck is the quality of the render: a 4096x4096 render like Figure~\ref{fig:Estra_AVL_render} took $\sim$8 hours on an HP z820 workstation with a single Xeon E5-2670, 2.5 GHz 10-core processor. If the number of particles used was reduced, or the pscale threshold and/or image quality was reduced, this render would be less computationally expensive. \par


\subsection{Finding Physically Interpretable Clusters in the Synestia}

One of the key aims of this work is to find and visualize ``physically interpretable'' clusters, whereby each cluster has a corresponding physical meaning or structure to automate and guide the visualization. With the color-mapping process designed to highlight individual clusters, physical interpretability is a requirement for the resulting visualizations to be meaningful, and not arbitrarily representations of the data which cannot be soundly analyzed. \par

In order to understand which clusters are physically meaningful, our 2D phase space cluster results were imported into Houdini's Scene View. In \texttt{Estra}, every particle is assigned a cluster membership, and this cluster membership ID becomes a new data attribute. Each particle retains its unique cluster assignment, and from there the shader can be extended to assign a unique color to each cluster. To do so, one can either create a new color ramp or copy an existing one, set its interpolation to `Constant', and map discernible colors to match the number of clusters. Once the clustering assignment ID is mapped to a color, each particle will be colored by its appropriate grouping. With Houdini's node network, all one has to do is change the connection to any one of the color ramps to achieve the desired render. \par

In this example, each cluster color pairing (Cluster 1--Gold, Cluster 2--Magenta, etc.) as shown in Figure~\ref{fig:5_clus_GMM_result} is retained in the shader. Each particle's 2D phase space cluster assignment manifests in 3D position space, from which the wide array of Houdini tools can be used for data exploration---rotating about the data, taking slices, isolating one cluster, etc. Seeing and interacting with the cluster-colored data in 3D position space, combined with the knowledge of the simulated data, enables the researcher to determine if the resultant clustering assignments have physical meaning, or ``interpretability''. If the result appears nonsensical or arbitrary, likely another clustering algorithm should be used and further explored. \par

\begin{figure*}
\centering
\includegraphics[width=16cm]{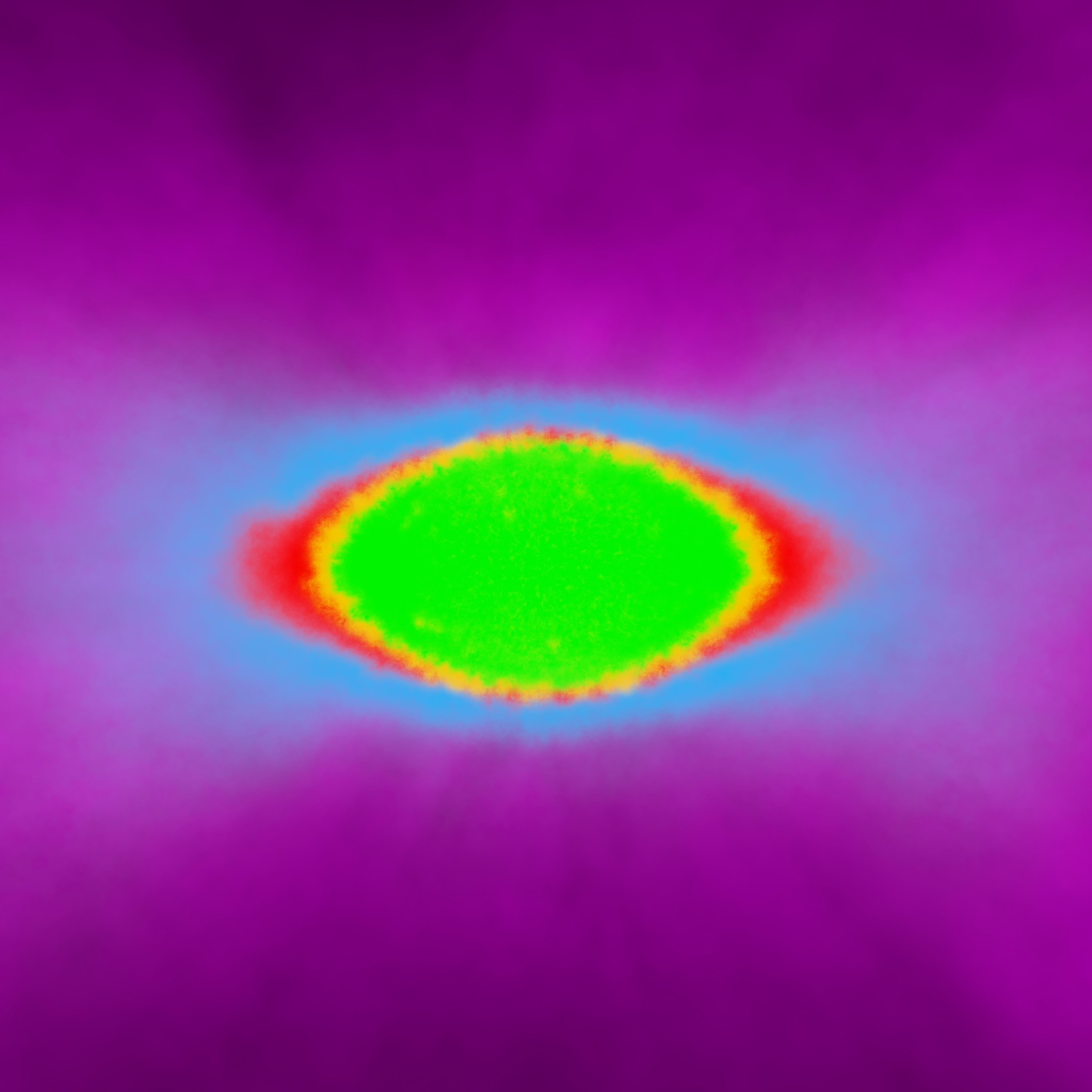}
\caption{A false-colormap render a cut-through of the far half of the synestia to emphasize the physical manifestation of the clustering results. The colors shown here match to those of temperature-entropy phase space plot (Figure \ref{fig:5_clus_GMM_result}) for easy comparison. The green region is the ``lower mantle'' of the synestia; yellow is the ``transition region'' containing a rapid increase in entropy with depth; red is the ``supercritical region'' of highly shocked silicate material which typically has specific entropies greater than the critical point entropy and so is either supercritical fluid or high-pressure vapor; cyan is the ``isentropic pure-vapor region'', because its constituent particles were forced to be isentropic during post-processing; and magenta is the outer ``vapor dome region'', where particles lie along the vapor side of the liquid-vapor phase boundary. This is rendered with the same custom shader as in Figure~\ref{fig:Estra_AVL_render}(a), with the only difference being the temperature ramp, which here maps by cluster ID and not by assigning color to a temperature transfer function from GMM results.}
\label{fig:slice_clus}
\end{figure*}

For our example post-impact synestia, the 5-cluster GMM result is shown in Figure~\ref{fig:slice_clus} (with earlier examples already displayed in Figures~\ref{fig:Scene_View_GMM_5_clus}, ~\ref{fig:GMM_5_cluster_outlier}). It is rendered in Houdini with a false colormap identical to the color scheme used in Figure~\ref{fig:5_clus_GMM_result}. This render is oriented such that the rotation axis points upwards through the central region. Also, to clearly see the physical meaning of the clusters, this render is a cut-through of the far half of the dataset, such that particles on the near side of the camera (our current view) do not occlude those on the far side along the line of sight through the synestia. In other words, only the far-half of the synestia is rendered, enabling us to investigate its inner layers via our line of sight. \par 

A motivating factor for choosing the 5-cluster GMM result is that it best clusters the data into its constituent components: the green region is the ``lower mantle'' of the synestia which experiences only moderate heating during the impact and so is still of relatively low entropy; red is the ``supercritical region'' of highly shocked silicate material which typically has specific entropies greater than the critical point entropy and so is either supercritical fluid or high-pressure vapor; yellow is the ``transition region'' between the supercritial fluid (red) and lower mantle regions (green); cyan is the ``isentropic pure-vapor region'', because its constituent particles were forced to be isentropic during post-processing (see \cite{Lock18}); and magenta is the region where particles lie along the vapor side of the liquid-vapor phase boundary in the outer ``vapor dome'' region. For the first time, the different components are easily discernible and can be understood in their proper context. It is important to remember that 1) these associations are interpretations by those who created the simulation, and 2) this information was not used to inform the clustering algorithm beforehand as \textit{a priori} knowledge. \par

\begin{figure*}
\centering
\includegraphics[width=18cm]{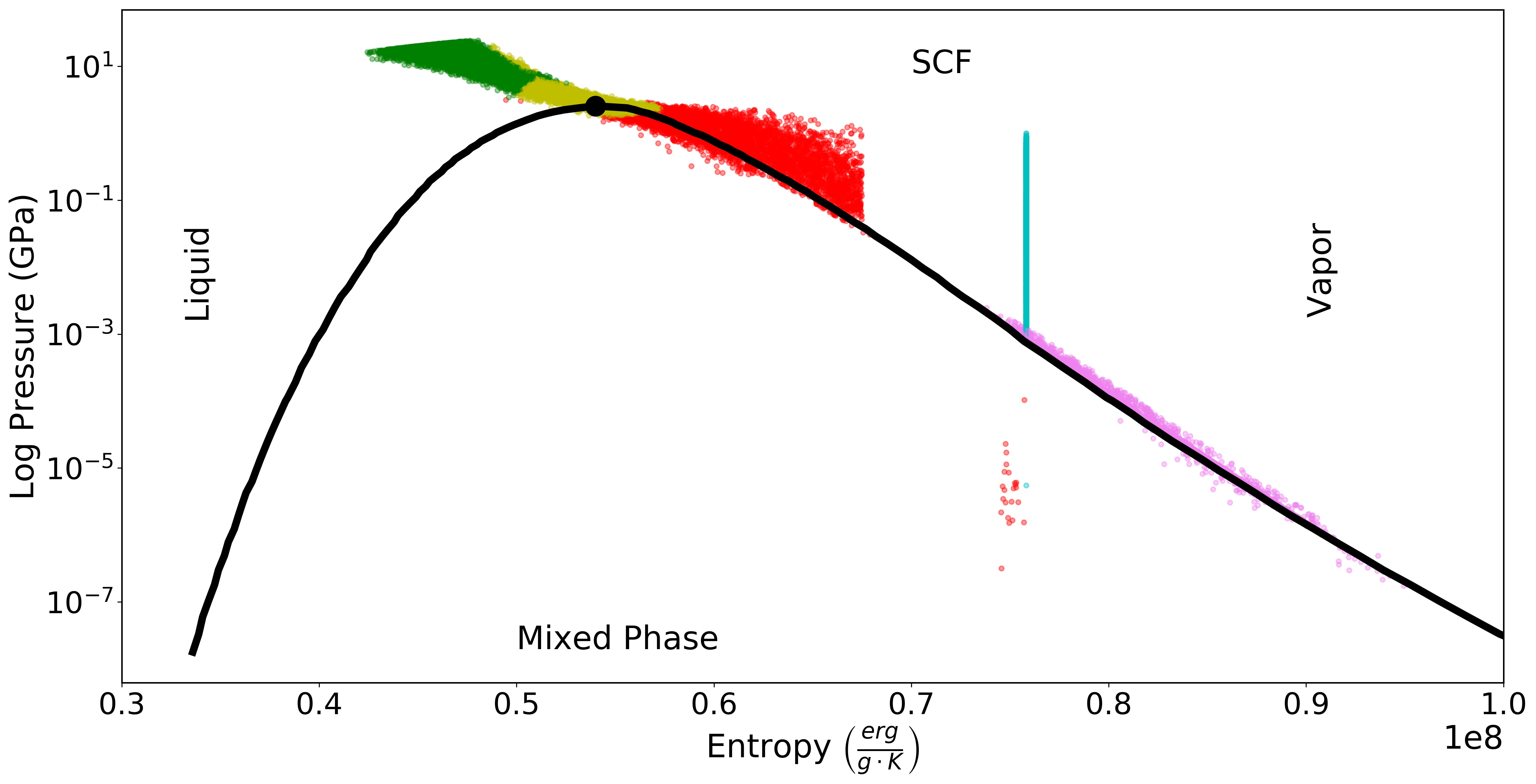}
\caption{Our 5-cluster GMM results plotted in a specific entropy-pressure phase space, where cluster membership is the same as in Figure~\ref{fig:5_clus_GMM_result}. In this phase-space, the liquid-vapor phase boundary is a dome-shaped
curve (black line). The black dot on the vapor dome is the critical point for the equation of state used in these
simulations ($S_{crit}$ = 5.40e7 erg K$^{-1}$ g$^{-1}$, $p_{crit}$ = 2.55 GPa, $T_{crit}$ = 8,810 K, $\rho_{crit}$ = 1.68 g cm$^{-3}$). Material to the left of the dome is liquid, material to the right of the dome and below the critical point is vapor, material above and to the right of the critical point is supercritical fluid (SCF), and material within the dome is a mixture of both liquid and vapor.}
\label{fig:S_logP}
\end{figure*}

As an additional check to confirm our structure analysis, we plotted the 5-cluster GMM results against the liquid-vapor phase boundary in pressure-specific entropy space, shown in Figure~\ref{fig:S_logP}.
The yellow transition region between the green lower mantle and the red supercritical region contains the critical point. The critical point marks the transition between liquid and supercritical fluid/liquid, a distinct change in the thermodynamic properties of the silicate, and so has been correctly identified by the clustering algorithm as a key region connecting, yet distinct from, the inner liquid and supercritical regions. Our clustering interpretation has also identified the region that constitutes supercritical fluid or high-pressure vapor as a distinct group, and the magenta outer vapor dome region neatly incorporates the particles that lie on the vapor side of the liquid-vapor phase boundary. Additionally, the cyan isentropic pure-vapor region had been clearly separated from the rest of the structure and is distinct from the magenta vapor dome cluster, thus confirming that the Kmeans clustering algorithm, which combined the pure-vapor and isentropic clusters (see Figure~\ref{fig:5_clus_Kmeans_ent_rho_result}), would indeed have been a poor choice. Our 5-cluster GMM algorithm distinctly separates physically significant regions within the synestia. \par

\begin{figure}
\centering
\includegraphics[width=\columnwidth]{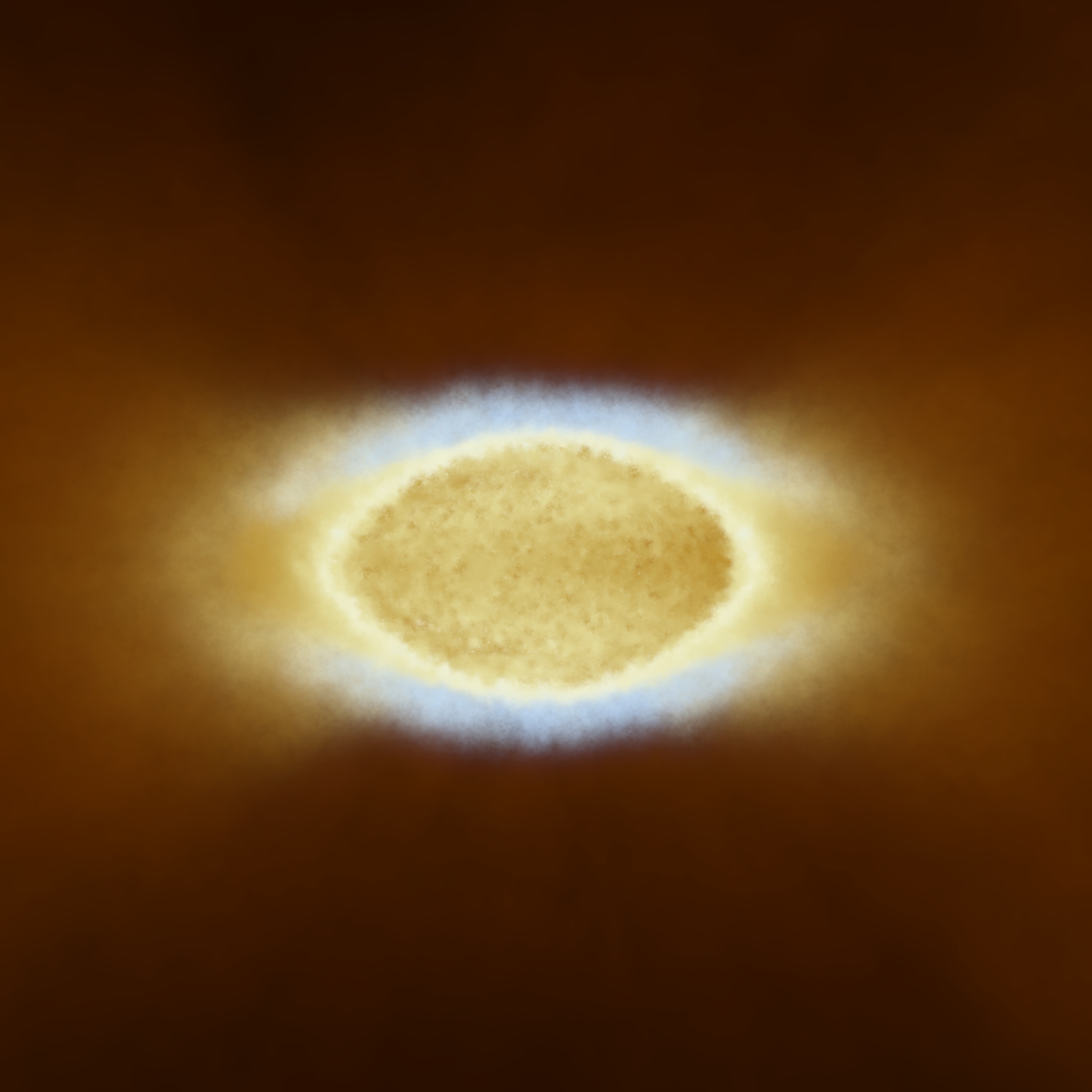}
\caption{Same as Figure~\ref{fig:slice_clus}, but rendered with both the non-uniform, emissive shader as in Figure~\ref{fig:Estra_AVL_render}(a) and the same temperature ramp as in Figure~\ref{fig:temperature_ramp}(a). We can clearly see some individual physical components of the synestia. From this, the hottest component of the synestia (light blue)---the isentropic region---is the outer surface layer perpendicular to the rotation axis, and not the innermost central region. Similar views can help scientists understand how their data is structured, and how different parameters map to those structures.}
\label{fig:slice_color}
\end{figure}

Retaining the same view as Figure~\ref{fig:slice_clus}, we can re-render a scene with both the non-uniform, emissive shader as in Figure~\ref{fig:Estra_AVL_render}(a) and the same temperature ramp as in Figure~\ref{fig:temperature_ramp}(a), shown in Figure~\ref{fig:slice_color}. 
Here, we can nearly discern the five structures as they appear in their individually colored cluster assignment. The innermost lower mantle is slightly cooler (more brown) than the transition region and supercritical region. This makes sense, as the supercritical region contains more highly shocked material that that in the lower mantle, and is of a higher temperature (yellow/white). The isentropic region contains the hottest material (most blue) on the outer surface layer perpendicular to the rotation axis material, which gets colder as the material occupies the outer edges of the wings. This is evident from the isentropic feature in Figure~\ref{fig:5_clus_GMM_result}. However, because of this temperature change, it is hard to tell to what extent this isentropic feature is due to this emissive shader, unless one looks at Figure~\ref{fig:slice_color} for reference. \par 

It is important to note that the stark change in temperature from the moderate, light yellow transition region to the hot, blue isentropic region is an artificially enforced boundary due to the post-processing of data by the scientists who created the simulation. This post-processing is designed to simulate thermal equilibrium of the post-impact body by processes not captured in the original SPH code but has the unfortunate consequence of introducing an unrealistically sharp boundary in the structure. In actuality the transition between these layers will be more gradiated. An advantage to a scientist visualizing their own work is that they will know artificialities such as this and can take steps to reduce their prominance in visualizations. \par 

The coolest material (dark brown)---which also has the highest specific entropy of silicate material---lies along the vapor-phase boundary in the outer edges of the bulge and disk. This, too, makes physical sense, as although the outer layers contain highly shocked, high-entropy material they are at low pressures where the temperature on the liquid-vapor phase boundary is relatively low. \par

\begin{figure}
\centering
\includegraphics[width=\columnwidth]{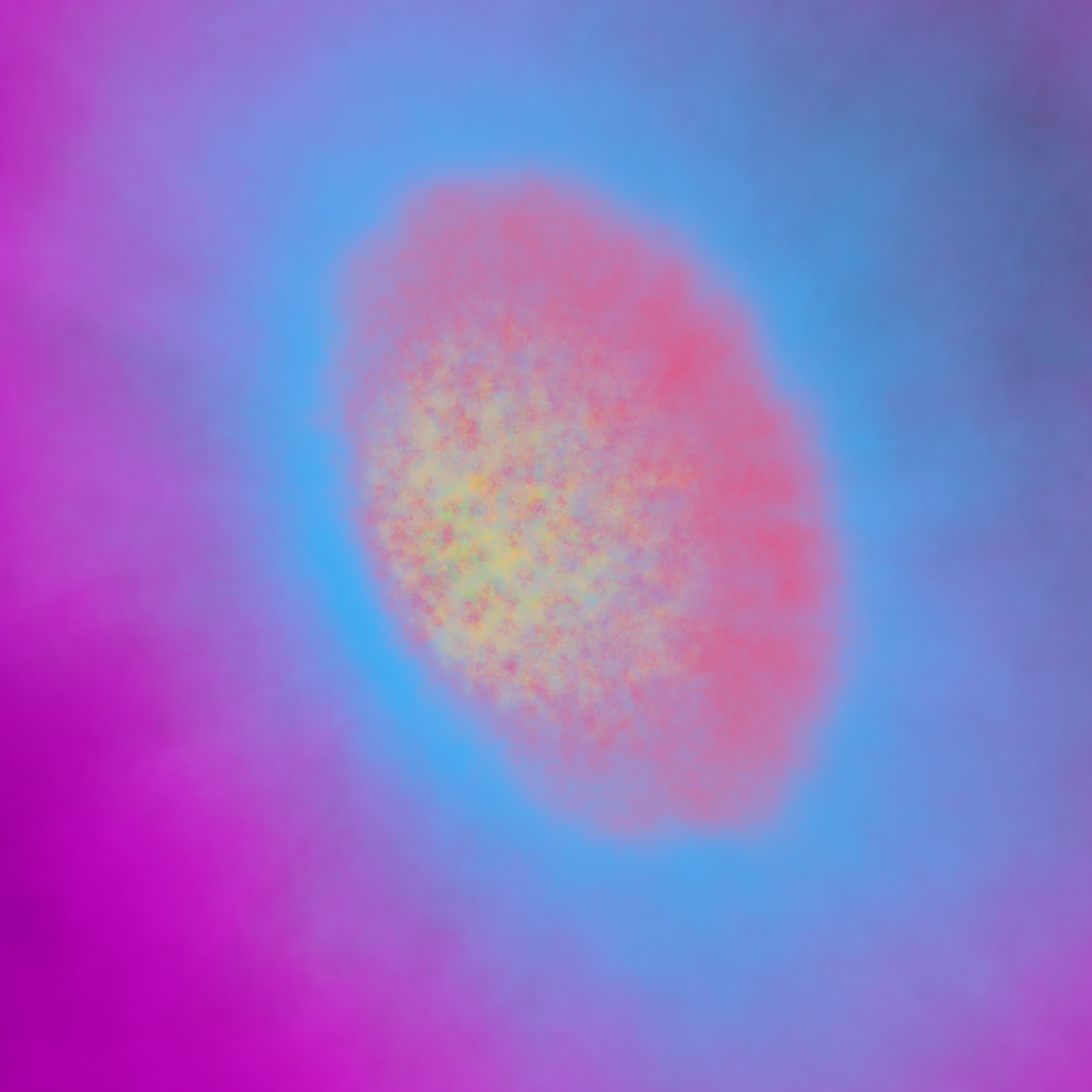}
\caption{Same as Figure~\ref{fig:Estra_AVL_render}(a), except the colormap is discrete and colored by the same 5-cluster GMM assignment as in Figures~\ref{fig:5_clus_GMM_result},\ref{fig:slice_clus}.}
\label{fig:map_id_clus_final}
\end{figure}

Similarly, we can render the same view as in Figure~\ref{fig:Estra_AVL_render}(a) but shading with the cluster assignment colors from the GMM, seen in Figure~\ref{fig:map_id_clus_final}.
From this angle (no slicing involved), it appears that the dusty ring of material is dominated by material from the supercritical region cluster. The brightest bulge material originates from the hottest isentropic particles. This is most apparent in Figure~\ref{fig:Estra_AVL_render}(a) as opposed to Figure~\ref{fig:Estra_AVL_render}(b), as intended. The spotty/patchy appearance of different temperature material in the bulge center in this line of sight is due to our view containing contributions from several different layers of the synestia. \par


\subsection{Validating the Visualizations}\label{subsec:validation}

Validating cinematic visualizations is a long-standing problem in the field, in essence due to its inherent nexus of art (visual effects) and science. Particularly, it is challenging to accurately represent the data in a way that is understandable to experts and non-experts alike \citep{Borkiewicz17}. Even in the case of this work, where the purpose of the visualization is tailored towards experts, how does one quantify a validation metric to ascribe its ``goodness''? When is the visualization ``correct'' enough to where it can be published in an academic journal or consumed by a general audience? With a medium that mixes qualitative and quantitative regimes, there is still no singular answer, but here we discuss several conventions. \par

Whilst creating the visualization, visualization teams consult peers and domain scientists---especially those responsible for producing the data---frequently for feedback. This includes, but is not limited to, explaining the purpose of the simulation and establishing its context in the field at large; checking factual accuracy on a documentary script; rerunning, fixing, or filling in gaps of the simulation; and, most importantly, vetting that the visualization does not convey essential aspects of the simulation that is objectively wrong \citep{Borkiewicz17, Borkiewicz19a}. It is up to the experts involved in both the visualization and science communities to agree that the visualization is an accurate (or accurate as possible) representation of the data. However, what ``accurate'' is in this context depends on the nature or purpose of the work. Visualizations are not a complete and holistic depiction of reality; some aspects are presented at the expense of others. Thus, the criterion for validity is not an absolute accuracy, but whether the visualization communicates some points effectively, or clarifies and reveals relationships between variables or physical processes, etc. Subsequently, this implies that accuracy or validity also depends on the purpose of the work. \par

Despite not being a concretely defined metric, accuracy is an important requirement because it is impossible to guess in the future who will see the visualization and in what context, and subsequently be misinformed. In fact, people psychologically tend to believe visualizations are true \citep{Borkiewicz17, Borkiewicz19a}, and thus if the final visualization conveys a biased or incorrect
message---whether intentionally or unintentionally, even if based on unbiased data---audiences can fall prey to scientific misconceptions\footnote{For a holistic overview of how and why people become misinformed about scientific concepts, see \cite{Scheufele2019} \& references therein.}.  \par

When the cinematic visualization is designed for consumption by general audiences, determining its success is harder. Audience testing involves entrance and exit interviews or questionnaires, and are meant to document audience reactions as opposed to learned intuition \citep{Borkiewicz17, Borkiewicz19a}. As expected, assessing the visualizations' success is not easily quantifiable. Further complications arise when animations of highly specialized dynamic subject matter are visually complex, which can negatively affect the learning experience \citep{Lowe03}.  \par

A third metric is an ``eye test'', where different representations/colors/camera positions, etc. of the same dataset are shown to different audiences, but all of whom are given the same questionnaire. Although this cannot establish the baseline of how ``good'' a visualization is, it can inform which visualization is ``better'' than another in whatever its particular usecase. \par

In the case of this work, the cinematic synestia visualizations are for field experts. 
Further, our clustering-informed visualizations have a tractable interpretation of its phase-space clusters, put into proper context of the physical structures and processes at play, thus completing the goal set by this work. This is as close to a metric of ``accuracy'' we can achieve. Future studies would benefit from some quantitative metric by which one could ascribe a ``goodness rating'' to the final visualizations, but that is beyond the scope of this work. \par

\section{Conclusion}\label{sec:Conclusion}

We have demonstrated the feasibility of using the visual effects software Houdini for cinematic astrophysical data visualization, informed by machine learning clustering algorithms. We outline a step-by-step process from a raw simulation into a finished render that can be utilized by non-experts in the field of visualization to achieve production-quality outputs. We used machine learning clustering algorithms and simple assumptions to inform the visualization process via our python pipeline \texttt{Estra}. As proof of concept, we used a single timestep of a post-impact, thermally-equilibrated Moon-forming synestia from \cite{Lock18}. \par

We showed the results of a 5-cluster GMM algorithm, which clustered the data into five distinct, physically-meaningful or ``interpretable'' clusters: a lower mantle region, transition region, supercritical region, isentropic pure-vapor region, and outer vapor dome region. By having the minimum, mean, and maximum values of these clusters in temperature-entropy phase space inform the colormap, we are able to emphasize these distinct structures in the render. Moreover, by assigning each cluster membership as an attribute and read into Houdini, each 2D phase-space cluster can be displayed in 3D position space. This will enable any researcher to better understand their data and better interpret the clustering results, particularly with Houdini's wide array of data-handling tools. \par

We rendered visualizations of the synestia with a shading network informed by the 5-cluster GMM result, and compared this to an identical render with a custom shading network by the Advanced Visualization Lab at the National Center for Supercomputing Applications for ``Imagine the Moon''. We showed that with simple assumptions and the clustering result, it is possible to achieve a render similar in quality to a professional team of visualization designers. Furthermore, using our clustering and visualization pipeline, other scientifically-informed renders (e.g. segment and show distinct, meaningful clusters) can be compared in a fraction of the time. Renders simply of the particles' clustering assignment can help understand the context of the different physical structures and inform scientific discovery. \par

Our results are significant in that they help realize the context of 2D phase-space information in 3D position space in a relatively simple manner. Because Houdini is a visual effects software, it can provide benefits not found in scientific software such as (but not limited to) excellent camera and animation controls, general-purpose ability, high-quality renders, robustness from large user and developer base, etc. Though, Houdini will be best paired in tandem with scientific software due to their native ability to read specific data formats, and there have been custom software that best utilize the two for visualizations such as \texttt{ytini} \citep{Naiman17}.

Cinematic astrophysical data visualization is an underdeveloped field in the literature, and its practices are not widely adopted by the astronomical community. By establishing such a literature, we hope to equip astronomers with the tools, skills, and knowledge to develop their own visualizations for publications, public outreach, prototype testing, etc.  \par

We suggest future endeavors focus on developing tools for multi-timestep data, where cluster assignments are made temporally. Specifically, tools that track and visualize particles by their cluster assignment in each time step would allow researchers to see not only how clusters change and evolve over time, but how their members do as well. This is invaluable information, and can lead to a better understanding of physical processes acting on both small and large scales. Lastly, future work which incorporates extrinsic audience perceptual metrics can help visualization designers better understand how factors such as lighting, occlusion, and color influence affect an audience's perception of the spatial and scientific reality of a dataset. \par

\section*{Acknowledgements}

The authors would like to thank all the additional members of the Advanced Visualization Lab at the National Center for Supercomputing Applications for their invaluable support, input, and insight: Jeffrey Carpenter and Colter Wehmeier. We also thank Matthew J. Turk for his vital role in overseeing many aspects of this work. Additionally, this research was made possible, in part, by funds from the 2019 Fiddler Innovation Fellowship generously supported by Jerry Fiddler. SJL acknowledges funding from the Caltech Division of Geological and Planetary Sciences and NSF through grants EAR-1947614 and EAR-1725349.




\bibliographystyle{mnras}
\bibliography{Estra_ref} 




\appendix

\section{Other Clustering Algorithms}

Because the choice and interpretation of various clustering algorithms are vital to producing structurally-meaningful results, we provide a quick overview of several widely-used methods.

\subsection{Kmeans}

The first of many unsupervised clustering algorithms available to use in \texttt{Estra} is Kmeans\footnote{\url{https://scikit-learn.org/stable/modules/generated/sklearn.cluster.KMeans.html}}, through the \texttt{sklearn.cluster} module. As a whole, Kmeans is likely the most popular of the array of the clustering methods, perhaps due to its scalability to large sample sizes. After specifying a desired number of $n$-clusters (which itself is either a known prior, or an estimated one) the data will be split into $n$-groups of equal variance by minimizing the inertia (also known as ``within-cluster sum-of-squares'') criterion. More specifically, it is the mean values of each cluster, referred to as centroids, that minimize the inertia. A minimization of the inertia is optimizing how internally coherent the clusters are, such that each member within a particular cluster is most similar to its members and most dissimilar to members outside that particular cluster. Thus, generally speaking, Kmeans is useful for relatively few number of clusters with approximately even cluster sizes of flat geometry, where no explicit structure relates one cluster to another. \par

The Kmeans cost function is given by

\begin{equation}
    \sum_{i=0}^{n} \min_{\mu_j \in C}(||x_i-\mu_j||^2),
\end{equation}

where the standard $L_2$ Euclidean distance is minimized between each $i^{th}$ cluster data point $x_i$ and the collection of centroids $\mu_j$ in the set $C$. Once the algorithm satisfies some stopping criterion (say, some maximum number of iterations is completed or no data points change their cluster assignment between iterations), the output will be dataset labels of all data points for user-selected K-clusters. This convergence may be a local optimum, as some randomized starting centroids will ultimately provide better results than others. We suggest running this algorithm multiple times and seeding with different initial guesses. \par

As with any algorithm, there are downsides to Kmeans and situations in which other algorithms are better suited. One such drawback is for data with irregularly shaped manifolds, because a critical underlying assumption of the inertia criterion is that the clusters are convex and isotropic. Although the algorithm will reach a convergence, it is unlikely then that this particular result will correspond to a physical meaning; it will more likely be a mathematical artifact. Another case where Kmeans suffers is in reconciling its Euclidean distance measurements when higher dimensions are involved---namely the ``curse of dimensionality''. When plotting very high dimensional data in a lower dimensional space, the distances between points start to lose meaning; by nature of not being able to aptly express the true nature of the data in all of its dimensionality, some variance is lost. Thus, as Kmeans utilizes standard Euclidean distance measurements, these distances become inflated in higher-dimensional space. In this case, then, we suggest first running some dimensionality reduction such as Principal Component Analysis (PCA), Random Forest (RF), t-Distributed Stochastic Neighbor Embedding (t-SNE), Non-negative Matrix Factorization (NMF), etc. \par

\subsection{DBSCAN}

DBSCAN\footnote{\url{https://scikit-learn.org/stable/modules/generated/sklearn.cluster.DBSCAN.html}} (Density-Based Spatial Clustering of Applications with Noise) is quite different from Kmeans and other clustering algorithms in that it does not care about the shape of the clusters \citep{Ester96}. Instead, the main focus of the algorithm is to separate areas of high density from low density. DBSCAN achieves this by using two main parameters: $\epsilon$(eps) and $min\_samples$ (MinPts in \cite{Ester96}). The $\epsilon$ parameter is used to assign a maximum distance for which two points can be considered to be within the same neighborhood, and the $min\_samples$ parameter is the number of datapoints needed in a neighborhood for a datapoint to be labeled a ``core point''. Put another way, it assigns a core point $p$ if there are $\geq$ $min\_samples$ points within a given distance $\epsilon$ of $p$ itself. Then, if another point $q$ is contained within a distance $\epsilon$ of $p$, the point $q$ is known to be ``directly reachable'' to be $p$ (where ``directly reachable'' only applies to core points). Only points directly reachable to a core point, including both non-core and the core points themselves, compose a single cluster; all points that fall outside of the range for which any one core point is directly reachable is deemed an ``outlier'' or ``noise point'', and is not considered part of any cluster. Note that the non-core points directly reachable to only $\leq$ $min\_samples$ core points are called ``edge points''. However, there is one more aspect in deciding which points belong to which cluster: ``density connectedness''. If there is another point $o$ such that any two points $p$, $q$ are both directly reachable from it, the points $p$ and $q$ are ``density-connected''. Thus for any particular cluster, all points contained within it are 1) mutually density-connected and 2) directly reachable to at least one core point (making this a region of ``higher density''). \par

As a density-based clustering algorithm, DBSCAN offers some critical advantages over Kmeans and various traditional clustering methods. One such benefit is that the number of clusters is not required to be known \textit{a priori}; instead, a knowledge of the dataset on how to best set the values for $\epsilon$ and $min\_samples$ is suggested, but not required. Further, as aforementioned, it can find arbitrarily shaped clusters, even those which are not linearly separable (non-flat geometry), in addition to finding unevenly sized clusters, as long as they are of the same relative density. Inherent to its formulation, DBSCAN is robust to outliers as well, as it factors in noise into its cluster finding process. \par

The main disadvantage of DBSCAN is in regard to its memory consumption when large data samples are involved. In fact, the \texttt{sklearn.cluster} Python package for DBSCAN involves the worst-case memory storage of $\mathcal{O}(n^2)$ floats when a full pairwise similarity matrix is used\footnote{See DBSCAN's user guide in scikit-learn's documentation.}. The pairwise similarity matrix is used in the case where its default method of $k$d-trees or ball-trees are not applicable, such as for sparse matrices; otherwise, the user can choose their preferred nearest neighbors module to compute pointwise distances and find nearest neighbors. Another main disadvantage lies with the curse of dimensionality, as before. DBSCAN uses $k$d-trees or ball-trees for its nearest neighbors search, and in high dimensional space, these methods (particularly $k$d-trees) are not well suited for efficiently finding such nearest neighbors. Often times a simple exhaustive search is as efficient, and it is better to use more approximate nearest neighbor methods. Further, as pointed out above, higher dimensions render the Euclidean distance metric for all practical purposes useless, and DBSCAN uses Euclidean distance as a measure for determining which points are ``directly reachable'' to each other. And lastly, if there is a wide range of densities within the data, the particular chosen values of $\epsilon$ and $min\_samples$ may not be appropriate when scaled to all clusters. 

\subsection{Variational Bayesian Gaussian Mixture Model}

Variational Bayesian Gaussian Mixture Model is a variant of GMM that utilizes variational (Bayesian) inference \citep{Jordan99, Wainwright08}, instead of EM, to fit the model \citep{Attias00}. This is known as variational Bayesian estimation of a Gaussian mixture model, a form of approximate posterior inference. \par

We desire to know the true posterior of the distribution, which from Bayes' Rule is  
\begin{equation}
    p(\textbf{z}|\textbf{x},\alpha) = \frac{p(\textbf{z},\textbf{x}|\alpha)}{\int_Z p(\textbf{z},\textbf{x}|\alpha)} = \frac{p(\textbf{z})p(\textbf{x}|\textbf{z})}{p(\textbf{x})},
    \label{eq:BayesRule}
\end{equation}
where $\textbf{z}$ = $z_{1:m}$ are our hidden, or latent, variables, $\textbf{x}$ = $x_{1:n}$ are our observations, or data (e.g. the particles of the simulation), and $\alpha$ are the fixed additional parameters (though, if the parameters are lumped into $\textbf{z}$, then $\alpha$ becomes the hyperparameters). Computing the posterior distribution is what is known to be an inference problem. \par

Finding the true posterior, or, as will be the case, finding an approximate posterior, is important for predictive distribution (given the data $\textbf{x}$, compute the conditional probability of a new observation $x^*$, denoted as $p(x^*|\textbf{x})$), finding modes, investigating the posterior over hidden variables, etc. Unfortunately, computing the normalization constant of the posterior for many complex models, including Gaussian mixture, is often analytically intractable. One could use Markov-Chain Monte Carlo (MCMC) sampling \citep{Hastings70, Gelfand90} to find the true posterior, but with many parameters, convergence can be prohibitively slow, and it does not scale to large data as well as variational inference. \par

A simpler, tractable family of distributions $\mathcal{D}$ over the latent variables $\textbf{z}$ is formed with its own collection of variational parameters $\nu$, i.e. $q(\textbf{z}|\nu)$. These variational parameters are chosen so that $q$ becomes a proxy for the desired posterior. In essence, $p(\textbf{z}|\textbf{x}, \alpha) \approx q(\textbf{z}|\nu, \alpha)$. To achieve this, one finds the particular family member that minimizes the Kullback-Leibler (KL) divergence with respect to the exact posterior 
\begin{equation}
    q^*(\textbf{z}|\nu, \alpha) = \operatorname*{argmin}_{q(\textbf{z}|\nu, \alpha) \in \mathcal{D}} \operatorname*{KL}(q(\textbf{z}|\nu, \alpha)||p(\textbf{z}|\textbf{x}, \alpha)),
\end{equation}
where $q^*(\cdot)$ is the optimized member of each family, and KL is defined to be
\begin{equation}
    \operatorname*{KL}(q||p) = \int_z \operatorname*{log} \frac{q(z)}{p(z|x)} = \mathbb{E}\Big[\operatorname*{log}\frac{q(z)}{p(z|x)}\Big].
\end{equation}
Because the KL divergence is an asymmetric measure of the difference between two distributions $p$ and $q$, its minimization will ensure that $p$ is most similar to $q$. Thus, one can use the approximate conditional $q^*(\textbf{z}|\nu, \alpha)$ as a best guess to the posterior $p(\textbf{z}|\textbf{x}, \alpha)$, and subsequently replace the posterior with the approximation. In all, variational inference rewrites a standard inference problem as an $\emph{optimization}$ problem.
For a more rigorous mathematical justification and explanation, see \cite{Blei17}. \par

In the \texttt{BayesianGaussianMixture} class, two types of weighting schemes are available to be specified by the user: 1) a finite mixture model using Dirichlet Distribution, and 2) an infinite mixture model using Dirichlet Process (DP). The reason for implementing the Dirichlet distribution may be nebulous, so first consider the joint distribution $p(\textbf{z},\textbf{x})$. From Bayes' Rule, as shown in Equation \ref{eq:BayesRule}, the posterior is proportional to the product of the prior of the latent variable (cluster), $p(\textbf{z})$, with the conditional $p(\textbf{x}|\textbf{z})$. In a GMM, $z_i$ is drawn independent and identically distributed ($i.d.d.$) from the multinomial (or categorical) distribution, and the conditional $x_i|z_i$ is drawn from a Gaussian distribution. Because 
the Dirchlet distribution is the conjugate prior of the multinomial (and categorical) distribution, that means if the likelihood---the generative model---is multinomial (or categorical), then both the prior and the posterior are Dirichlet distributed. Then, with the form of the posterior known to be Dirichlet, it is easier to compute when using the variational Bayesian inference method. \par

From this point, all that is left is selecting between a finite mixture model with Dirichlet Distribution and a non-parametric infinite mixture model using Dirichlet Process (DP). DP (analogous to the Chinese Restaurant Process and Stick-breaking Process) is simply an ``infinite-dimensional'' Dirichlet defined by an infinite number of clusters and a concentration parameter. Although an infinite number of clusters are available only a finite number are used to generate the data. This enables the user to not have to pre-specify the number of clusters. In fact, this setting only requires the user to specify the concentration parameter and the upper-limit on the number of mixture components, and the value of the \texttt{weight\_concentration\_prior} will decide to use either all or only some of the components. The lower the value, the fewer components (some very close to zero); the higher the value, the more active mixture components (as well as more uniform). \par

\section{Renders with perceptually-uniform temperature colormaps}\label{subsec:nonu}

One of the decisions a visualization designer has to make when developing a visualization is what type of color map to use. A poorly chosen color map can trick human eyes into seeing non-existent features, mixing features, or missing features altogether. One such example is a rainbow color map \citep{Borland07, Moreland16}. A perceptually-uniform colormap takes human visual perception knowledge and gives readers a correct sense on the image intensities regardless of the display. The main motivating factor is biological---the human eye is more sensitive to brightness than hue \citep{Borkiewicz19a}. Thus, any subtle change in brightness is more readily recognized. For example, the blackbody colormap from \cite{Moreland16} has colors which are based on those from blackbody radiation, but are not exact according to its wavelength. Instead, they increase in brightness at a constant rate, and the luminance is perceptually linear (in CIELAB color space).

\begin{figure*}
 \begin{minipage}[c]{1.0\linewidth}
  \centering
  \begin{center}
  \textit{(a)}{\includegraphics[width=0.47\textwidth]{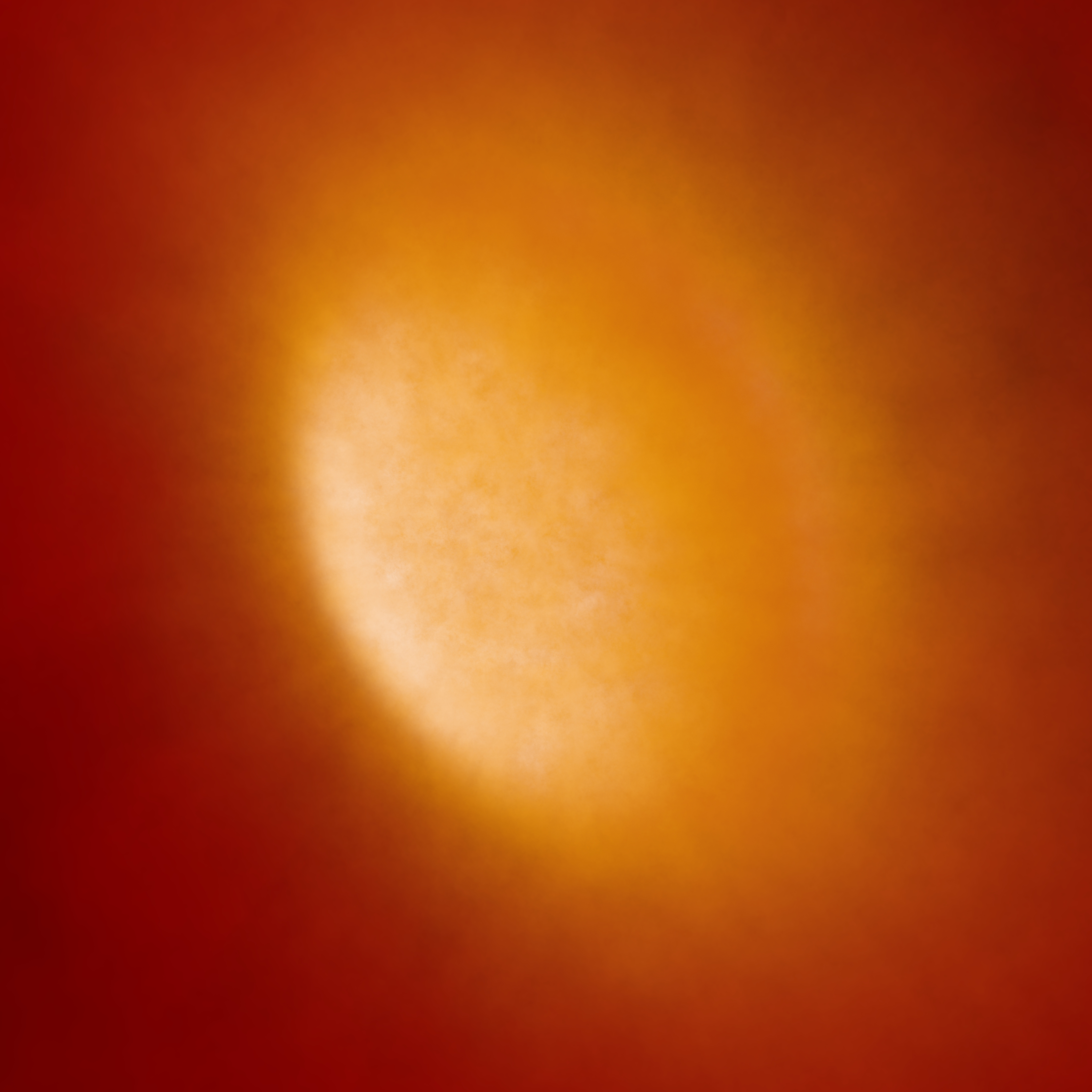}}
  \textit{(b)}{\includegraphics[width=0.47\textwidth]{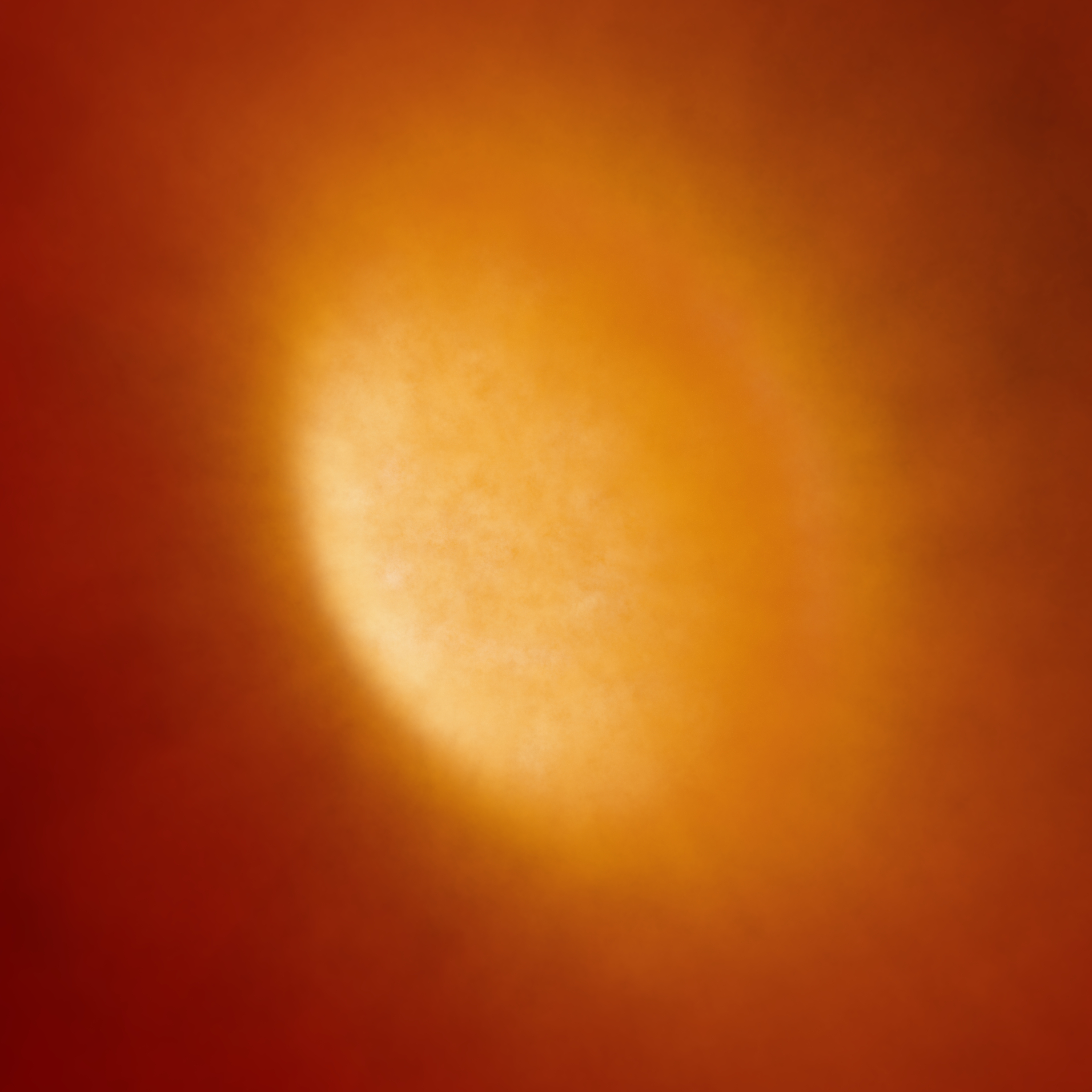}}
  \caption{Same as Figure~\ref{fig:Estra_AVL_render}, except this applies the same perceptually uniform colormap used in the black hole imaging by the Event Horizon Telescope: \textit{afmhot\_us} \citep{EHT19}. \textit{(a)} uses the 5-cluster GMM results with the simplified \texttt{Estra} shader network, and \textit{(b)} uses the custom AVL shader without clustering.}
  \label{fig:afmhot_us_renders}
  \end{center}
  \end{minipage}
\end{figure*}

Because the synestia is emissive, we rendered the image seen in Figure~\ref{fig:afmhot_us_renders} using a perceptually uniform colormap named \textit{afmhot\_us} from the \texttt{ehtplot}\footnote{\url{https://github.com/eventhorizontelescope/ehtplot}} library developed for \cite{EHT19}. 
This specific color map is symmetrized with linearity in lightness $J'$ as defined by the CAM02-UCS color appearance model introduced by \cite{Luo06}. A linearity of $J'$ values is a good approximation of uniform colormaps, and is the working definition of Perceptually Uniform Sequential colormaps by matplotlib\footnote{\url{https://matplotlib.org/tutorials/colors/colormaps.html}}. \par

As is evident, Figures~\ref{fig:afmhot_us_renders}(a),~\ref{fig:afmhot_us_renders}(b) are more gradual in their color change, and the dust ring appears to blend in more to its central region than those in Figure~\ref{fig:Estra_AVL_render}. Although interpretation is subjective, overall it is harder to distinguish here between the different structures of the synestia as seen in Figure~\ref{fig:map_id_clus_final}, compared to Figure~\ref{fig:Estra_AVL_render}(a). \par

The only appreciable difference between renders in Figure~\ref{fig:afmhot_us_renders}(a) and \ref{fig:afmhot_us_renders}(b) is the pink appearing highlights in the left side of the bulge, which marginally differentiates this isentropic component from the rest of the bulge. This is because in this example, the perceptually-uniform colormap makes it more difficult to discern the synestia in Figures~\ref{fig:afmhot_us_renders}(a),~\ref{fig:afmhot_us_renders}(b). Because of the linearity in brightness and a narrower color palette range, the color value of the bulge is more similar to the image background, making it more difficult to distinguish the demarcation between source and background. This is likewise for the dusty ring in the plane perpendicular to its rotation axis compared to the background. \par

CIELAB color space, a device-independent model which describes all colors visible to the human eye, has three related color spaces: CIE76, CIE94, and CIEDE2000. All three of these color spaces define a slightly different metric for color difference called $\Delta E^*_{00}$ \citep{Luo01}, which attempts to quantify how noticeable two colors are based on knowledge that the human eye is more sensitive to some colors than others. However, in all formulae of $\Delta E^*_{00}$, Figure~\ref{fig:Estra_AVL_render}(a) has approximately equal or greater values for all bulge-background and ring-background comparisons than Figures~\ref{fig:afmhot_us_renders}(a),~\ref{fig:afmhot_us_renders}(b). Because these formulae are based on Euclidean distance measurements in color space, the greater the value between two colors represents the greater the color difference, and thus the more perceptually different the two colors are. \par

For instance, an example pixel in the brightest part of the bulge (``reference'') and background (``sample'') of Figure~\ref{fig:Estra_AVL_render}(a) has $(R,G,B)=(209,214,192)$ and $(R,G,B)=(80,34,1)$, respectively\footnote{Before applying the formula, $(R,G,B)$ is converted to CIE-$(L^*, a^*, b^*)$ coordinates.}. According to the most recent CIEDE2000 definition\footnote{See \cite{Sharma04} for info on the CIEDE2000 color difference formulae and explanation.} given by
\begin{equation}
    {\Delta E^*_{00} = \sqrt{ \left(\frac{\Delta L'}{k_L S_L}\right)^2 + \left(\frac{\Delta C'}{k_C S_C}\right)^2 + \left(\frac{\Delta H'}{k_H S_H}\right)^2 + R_T \left(\frac{\Delta C'}{k_C S_C}\right) \left(\frac{\Delta H'}{k_H S_H}\right)}},
\end{equation}
we find $\Delta E^*_{00}=68.6$.
Meanwhile, Figure~\ref{fig:afmhot_us_renders}(b), has $(R,G,B)=(246,208,142)$ and $(R,G,B)=(136,46,12)$, respectively, which results in $\Delta E^*_{00}=51.7$. Hence, it is easier to discern the bulge and background in Figure~\ref{fig:Estra_AVL_render}(a) as opposed to in Figure~\ref{fig:afmhot_us_renders}(b). \par

Although using the clustering-informed results on a perceptually-uniform colormap doesn't appear to be easily distinguishable in this case, it does in Figure~\ref{fig:Estra_AVL_render}(a) where a non-perceptually uniform colormap is used.

In deciding the ``best'' render for this work, we argue that one which emphasizes by features of the inner regions of the terrestrial synestia from the extended disk-like regions is optimal. From the calculation of $\Delta E^*_{00}$, this is clearly Figure~\ref{fig:Estra_AVL_render}(a). However, each dataset is different, and a perceptually uniform colormap can be a quick and useful starting place for visualization teams or scientists. \par 



\bsp	
\label{lastpage}
\end{document}